# Effects of the Affordable Care Act Dependent Coverage Mandate on Health Insurance Coverage for Individuals in Same-Sex Couples ♣

Christopher S. Carpenter[1]

Gilbert Gonzales[2]

Tara McKay[3]

Dario Sansone[4]

This version: April 2020

**Abstract**
A large body of research documents that the 2010 dependent coverage mandate of the Affordable Care Act was responsible for significantly increasing health insurance coverage among young adults. No prior research has examined whether sexual minority young adults also benefitted from the dependent coverage mandate, despite previous studies showing lower health insurance coverage among sexual minorities and the fact that their higher likelihood of strained relationships with their parents might predict a lower ability to use parental coverage. Our estimates from the American Community Surveys using difference-in-differences and event study models show that men in same-sex couples age 21-25 were significantly more likely to have any health insurance after 2010 compared to the associated change for slightly older 27 to 31-year-old men in same-sex couples. This increase is concentrated among employer-sponsored insurance, and it is robust to permutations of time periods and age groups. Effects for women in same-sex couples and men in different-sex couples are smaller than the associated effects for men in same-sex couples. These findings confirm the broad effects of expanded dependent coverage and suggest that eliminating the federal dependent mandate could reduce health insurance coverage among young adult sexual minorities in same-sex couples.

♣ We thank seminar participants at Vanderbilt University for helpful comments. We gratefully acknowledge financial support from the Robert Wood Johnson Foundation Grant #76601. Results do not imply the endorsement of the Robert Wood Johnson Foundation or any other organization.

[1] Vanderbilt University, NBER and IZA. E-mail: christopher.s.carpenter@vanderbilt.edu
[2] Vanderbilt University. E-mail: gilbert.gonzales@vanderbilt.edu
[3] Vanderbilt University. E-mail: tara.mckay@vanderbilt.edu
[4] Vanderbilt University and University of Exeter. E-mail: dario.sansone@vanderbilt.edu

# 1. Introduction and motivation

Substantial research has documented that sexual minorities (lesbian women, gay men, bisexual individuals, and other non-heterosexual populations) have worse health outcomes, including increased prevalence of mental health and substance use disorders, HIV infection, and risk factors for chronic disease such as cigarette smoking and heavy alcohol consumption (Boehmer 2002; Bostwick et al. 2010; Carpenter and Sansone 2020; Cochran et al. 2013; Gonzales et al. 2016; Gonzales and Henning-Smith 2017; Gorman et al. 2015; Hatzenbuehler et al. 2008; Meyer 1995). Despite having greater health care needs, sexual minorities also experience barriers to medical care, as they are more likely to be uninsured and delay or forgo medical care because of financial cost (Buchmueller and Carpenter 2010; Dahlhamer et al. 2016; Gonzales and Blewett 2014; Heck et al. 2006; Ponce et al. 2010). These disparities have been identified and targeted for elimination by the National Academy of Medicine (IOM 2011) and the National Institutes of Health (Pérez-Stable 2016). Improving health insurance coverage and access to care may be one important lever for reducing sexual orientation-based disparities.

Prior research has examined how LGBTQ-specific policies - such as domestic partnership and same-sex marriage laws - impact private health insurance coverage for sexual minorities (Buchmueller and Carpenter 2012; Dillender 2015; Gonzales 2015), but very little research has examined the impacts of broad population-based health reforms on sexual minorities (Carpenter and Sansone 2020). The Affordable Care Act (ACA) represented one of the most important health insurance reforms in recent history, and a large body of research documents the effects of the ACA at reducing rates of uninsurance in the nonelderly adult population. In particular, the 2010 ACA dependent coverage mandate - which allows young adults up to age 26 to enroll as dependents on a parent's private health plan - significantly increased insurance coverage among young adults below age 26 compared to the associated change for slightly older individuals who were not eligible for parental coverage (Antwi et al. 2013; Barbaresco et al. 2015; Mulcahy et al. 2013; Sommers and Kronick 2012; Wallace and Sommers 2016).

In addition, numerous studies have examined the impact of the ACA dependent coverage mandate on racial and ethnic minorities (Chen et al. 2016; O'Hara and Brault 2013; Scott, Salim, et al. 2015; Shane and Ayyagari 2014), women (Robbins et al. 2015), rural populations (Look et al. 2017), and young adults with specific medical conditions and disabilities (Ali et al. 2016; Golberstein et al. 2015; Porterfield and Huang 2016; Saloner and Cook 2014; Scott, Rose, et al. 2015). To our knowledge, however, there is no research that has specifically examined the causal effects of the ACA dependent coverage mandate on sexual minorities.

In this paper we provide the first evidence on how the ACA dependent coverage mandate affected health insurance coverage for sexual minorities cohabiting in same-sex couples as well as how it affected disparities in health insurance coverage between same-sex couples and different-sex couples. There are several reasons to believe that the ACA dependent coverage mandate may have differentially affected health insurance coverage of sexual minority populations. First, sexual minority adolescents may be less able to take advantage of a parent's employer-sponsored health plan due to the higher likelihood of poor relationships with parents. A large literature in psychology and family development documents that discrimination and stigma surrounding the process of "coming out" can strain relationships between parents and sexual minority children (Cramer and Roach 1988; D'Augelli et al. 1998; Goldfried and Goldfried 2001; Heatherington and Lavner 2008; Radkowsky and Siegel 1997; Ryan et al. 2010; Savin-Williams 1989; Waldner and Magruder 1999). Sexual minority youth may receive less support and acceptance because of their sexual identity in early adulthood compared to heterosexual youth.[5] Some sexual minority individuals may even be disowned by their parents, as family rejection is a leading cause of homelessness among sexual minority youth (Durso and Gates 2012). Thus, strained familial ties would reduce the effectiveness of a dependent coverage mandate at increasing insurance for sexual minority young adults.

Second, sexual minorities may have fewer alternative sources of health insurance coverage than heterosexual individuals. The vast majority of adults in the United States obtain health insurance through their employer (Barnett and Vornovitsky 2016), and there is strong evidence that sexual minorities face potential barriers to employment, including labor market discrimination (Tilcsik 2011). Even for sexual minorities with employment, however, their same-sex partners and spouses may lack access to health insurance because historically employers have been less generous in offering insurance coverage to same-sex partners and spouses of employees than in offering insurance coverage to different-sex partners and spouses of employees.[6] Even in the presence of an employer offer of health insurance to a same-sex partner or spouse, an employed sexual minority individual with a same-sex partner or spouse may not feel comfortable effectively outing herself to her employer for fear of workplace reprisals, especially since most US states lack employment nondiscrimination protection on the basis of sexual orientation (MAP 2019). Thus, parental coverage may be an attractive source of insurance for sexual minority adults in same-sex couples, particularly for those without access to own employer-sponsored insurance.

Third, differences in health, human development, and socioeconomic status between sexual minorities and heterosexuals may result in differential demand or need for health insurance by sexual orientation. A large body of research shows that sexual minority adults are more likely to have college and advanced degrees compared to heterosexuals (Black et al. 2007; Carpenter and

---

[5] A 2013 Pew Research Center report indicated that, among a nationally representative sample of lesbian, gay, and bisexual Americans, the median age at which gay men told a close friend or a family member about their sexual orientation was 18; for lesbians the median age was 21 (Pew 2013). Our samples will focus on individuals in cohabiting same-sex romantic relationships, which is likely to be positively correlated with having come out to family members.

[6] The overwhelming majority of employers cover different-sex spouses under family insurance plans, and of course all individuals in different-sex couples always had the legal option to marry over our primary sample period (2008-2012). The same was not true for individuals in same-sex couples; nationwide access to legal same-sex marriage was only granted in the United States in 2015 in the United States Supreme Court ruling *Obergefell v. Hodges*, and employer surveys have shown that not all employers have adopted insurance benefits for legal same-sex spouses even after *Obergefell* (Dawson et al. 2016).

Gates 2008; Gonzales and Blewett 2014). If sexual minorities are disproportionately more likely to delay employment (where again the vast majority of Americans obtain health insurance) they may be more likely to need access to a parent's insurance plan. Relatedly, a range of health conditions and health behaviors prevalent among sexual minority adults may also influence the demand for dependent coverage. Sexual minority women, for example, are less likely to utilize family planning and contraceptive services as well as health care related to childbirth and labor (i.e. maternity care), and these are leading sources of insurance-related healthcare for heterosexual women in adulthood (Agénor et al. 2014; Agénor et al. 2017; Charlton et al. 2011; Charlton et al. 2014; Ela and Budnick 2017; Kerr et al. 2013; Tornello et al. 2014). On the other hand, sexual minority men may be more likely to need health care for conditions prevalent among this population, including sexually transmitted infections, smoking cessation, and substance use disorders (Gonzales et al. 2016; Green and Feinstein 2012; Institute of Medicine 2011; Wolitski and Fenton 2011). Thus, differential patterns in family relationships, employer behavior, human development, and health profiles will likely affect (in a direction difficult to predict *ex-ante*) the ability of a dependent coverage mandate to increase health insurance coverage for sexual minorities relative to heterosexual young adults.

Ultimately, whether the ACA dependent coverage mandate affected health insurance coverage of sexual minorities – and whether any such effects are different than the effects for heterosexual people – remains an empirical question. Using data from the American Community Survey (ACS), we provide the first evidence on this question by examining individuals in same-sex couples who were age-eligible for parental insurance coverage benefits (i.e., 21 to 25-year-old) before and after 2010 and comparing this difference to the associated difference for slightly older individuals in same-sex couples who were not age-eligible for the ACA dependent coverage provision (i.e., 27 to 31-year-old).

**2. The Affordable Care Act Dependent Coverage Provision**

The Affordable Care Act (ACA) was signed into law by President Barrack Obama in 2010, and expanded health insurance to millions of Americans through Medicaid expansions for low-income families and individuals and subsidies to purchase private health insurance for middle-income Americans. One of the first reforms to be implemented was the dependent coverage provision. Starting on September 23, 2010, this provision required employers to extend employer-sponsored health insurance to the dependent children of covered employees until 26 years of age.

Prior to the implementation of the ACA, more than 30 states enacted similar policies, but the impacts of state-level dependent coverage provisions were small (Cantor et al. 2012; Monheit et al. 2011). State-level dependent coverage provisions were often limited to a minority of employers that "fully insured" their employers through an insurance carrier (rather than "self-insured" employers). Numerous studies demonstrate that the federal dependent coverage provision had a relatively large impact on employer-sponsored insurance coverage, ranging between 6-8 percentage point increases in employer-sponsored insurance for young adults (Barbaresco et al.

2015; Cantor, Monheit, et al. 2012; Sommers and Kronick 2012). Unlike many of the pre-ACA state dependent coverage mandates, the ACA dependent coverage provision did not require that the dependent child be enrolled in school, did not require that the dependent be unmarried, and extended the age of dependency until age 26 (which was more generous than many states had implemented). As a result, it is not surprising that previous research has not found differential effects of the ACA dependent coverage provision among states with prior dependent coverage provisions when compared to the other states (Antwi et al. 2013; Barbaresco et al. 2015)

The dependent coverage provision of the ACA did not extend to spouses or unmarried partners of the policyholder's dependents, however. Thus, for individuals in same-sex and different-sex couples who we identify in the ACS, their only route to parental insurance coverage via the ACA was through the individual's *own* parent, not the parent of the spouse or partner.

## 3. Data

### 3.1 The American Community Survey

This study uses data from the American Community Survey (ACS) which is publicly available through IPUMS-USA at the University of Minnesota (Ruggles et al. 2020). The ACS is a nationally representative and repeated cross-sectional dataset. It contains demographic, economic, social, and housing information on 1% of the U.S. population (or approximately 3 million people each year). The large sample sizes available in the ACS facilitate studies on relatively small subpopulations, such as individuals in same-sex couples.

Importantly, the ACS has included a question on current health insurance status since 2008. We are able to identify whether the individual had any health insurance at the time of the survey, as well as the source of health insurance. Specifically, we can identify whether the individual had any of the following types: employer-sponsored insurance (ESI, including those covered by their employer, a spouse's employer, or another family member's current employer, former employer, or union), direct/privately purchased insurance, TRICARE (health insurance for active duty military personnel), Medicare, Medicaid, health care through the Department of Veterans Affairs (VA), or health care through the Indian Health Service. It is worth emphasizing that these categories are not mutually exclusive: individuals could be covered by more than one type of insurance (IPUMS 2019). We expect the ACA dependent mandate should primarily increase the likelihood that eligible young adults experienced an increase in employer-sponsored insurance. Unfortunately, the ACS does not ascertain whether a person with ESI was the policyholder or a dependent on a parent or a spouse's/partner's health plan.[7]

The ACS does not directly ask individuals about their sexual orientation. To identify a subset of sexual minorities, we follow a large body of prior research that uses intrahousehold relationships to identify individuals in same-sex couples (Black et al. 2000; Gonzales and Blewett 2014; Sansone 2019). Specifically, the ACS identifies a primary reference person, defined as "the person living or staying here in whose name this house or apartment is owned, being bought, or rented". For simplicity, we refer to the primary reference person as the household head. The ACS also collects information on the relationship to the household head for all members of the household, and the range of possible relationships includes husband, wife, and unmarried partner (as a different category than roommate). Notably, individuals of the same sex as the household head who describe their relationship to the household head as a 'spouse' were recoded to unmarried partners through 2012 in compliance with the federal Defense of Marriage Act (which did not recognize married same-sex couples for all federal purposes). Our final sample includes 2,781 and 3,614 men and women in same-sex couples, respectively, and 235,954 and 304,318 men and women in different-sex couples, respectively (all of whom are age 21-25 or 27-31).

---

[7] Other surveys contain this information (e.g., the Annual Social and Economic Supplement (ASEC) to the Current Population Survey), but we need the much larger sample sizes of the ACS to identify meaningful effects for sexual minorities.

### 3.2 Data quality and limitations

The ACS is a mandatory survey: although nobody has been prosecuted for not responding to the ACS survey (Selby, 2014), this approach significantly increases the response rate (typically above 90%) and data quality (U.S. Census 2017; U.S. Census 2019). Despite this, one key issue when dealing with same-sex couples is misclassification error: individuals can incorrectly report their sex or relationship to the household head. Since the proportion of different-sex couples is much larger than that of same-sex couples, there is the risk that several same-sex couples may actually be misidentified different-sex couples—even when such measurement errors may be rare. The U.S. Census Bureau implemented several changes between 2007 and 2008 to address this issue. These improvements resulted in a substantial drop in the reported number of same-sex couples between these two years, thus indicating more reliable estimates (U.S. Census, 2013).

Moreover, observations with imputed sex or relationship to the household head have been dropped to further reduce such measurement errors (Black et al., 2007; DeMaio et al. 2013; Gates and Steinberger 2007). It is also worth mentioning that older respondents in different-sex couples were the most likely to be misclassified as same-sex couples due to their lower levels of familiarity with the terminology pertaining to same-sex couples (Lewis et al., 2015). Since we focus on younger respondents, we exclude these cases by construction. Another advantage of ACS is that around a third of the households use Computer Assisted Telephone (CATI) or Personal Interviews (CAPI). In such interviews, respondents are asked to verify the sex of their same-sex husband/wife, thus reducing such miscoding (Gates and Steinberger, 2007).

Notwithstanding these issues, the U.S. Census and the ACS remain the largest and most reliable data on same-sex couples. For example, the across-metropolitan distribution of male same-sex couples in the 1990 Census lines up extremely well with AIDS deaths in 1990, a year during which AIDS deaths were predominately concentrated among gay men (Black et al., 2000). Fisher et al. (2018) found similar estimates when comparing economic statistics (such as income distribution) between Census and tax data. Using health data, Carpenter (2004) showed that individuals most likely to be in same-sex unmarried partnerships were indeed behaviorally gay, lesbian, or bisexual individuals, i.e. they exhibited sexual behaviors that were unlike those of individuals most likely to be in different-sex couples.

There are other surveys that contain information on sexual orientation or sexual behavior (e.g., the General Social Survey, or GSS). However, these alternative data sources have sample sizes that are too small for our analyses. The main disadvantage of using ACS data is that it is not possible to identify single LGBTQ individuals without a partner or same-sex couples who do not live together. Furthermore, since there is no individual-level information on sexual orientation, researchers cannot identify bisexual individuals in different-sex (or same-sex) couples (Hsieh and Liu 2019). In order to quantify these limitations, we have analyzed data from the 2013-2018 National Health Interview Survey (Blewett et al. 2020), which contain information on individual self-reported sexual orientation as well as household structure. Our calculations indicate that among 21-31 year old adults, 28 percent of self-identified sexual minority men (i.e., men who describe themselves as gay, bisexual, or 'something else') are in a household with a same-sex unmarried partner or same-sex spouse, while 39 percent of self-identified sexual minority women (i.e., women who describe themselves as lesbian, bisexual, or 'something else') are in a household with a same-sex unmarried partner or same-sex spouse. The associated share for self-identified heterosexual individuals is 47 percent. Thus, while the ACS same-sex couples are unlikely to represent the majority of sexual minority individuals in the United States, they do capture a substantial share (28-39 percent) of these populations of interest.

## 4. Econometric framework

We use a standard difference-in-differences approach to examine the impact of the ACA's dependent coverage mandate on young adults in same-sex and different-sex couples. Formally, the estimated difference-in-difference model is the following:

$$y_{igst} = \alpha + \beta(Treat_{ig} * Post_t) + \delta_s + \mu_t + \pi_g + x'_{st}\gamma_1 + x'_{igst}\gamma_2 + \varepsilon_{igst}$$

where $y_{igst}$ is whether individual $i$ in age group $g$ living in state $s$ at time $t$ had health insurance coverage. Our main outcome is whether an individual had any health insurance coverage, but we also analyze the other sources described above.

The coefficient of interest is $\beta$. $Treat_{ig}$ indicates whether an individual was in the treated age group 21-25[8] as opposed to the control group 27-31.[9] $Post_t$ indicates whether an individual was interviewed after or before 2010. Our main estimates focus on the years 2008-2012, but we also extend the time period up to 2018. Since the public use ACS does not include information on when during the calendar year the respondents were interviewed, and some insurers chose to comply with the ACA dependent coverage provision sooner than September 2010 (White House 2010), we exclude 2010 from most specifications since we cannot accurately determine treatment status. This also allows us to minimize the likelihood of anticipation effects, since it is possible that young people reduced their insurance coverage in the period between the enactment in March 2010 and the implementation of the reform in September 2010 (Antwi et al. 2013). Meanwhile, many employers updated their policies to allow young adults to enroll in the 2010 open enrollment periods for insurance that would begin the following year.

---

[8] We exclude individuals age 26 from the main analysis since we do not know if they were in the treatment group of the control group, though the vast majority of them were likely in the control group. As discussed in the empirical section, coding them as such does not materially change our findings. Strictly speaking, insurers were allowed to remove dependent children on the first day of the month following the month of the child's 26th birthday, although employers could decide to continue coverage for the whole calendar year beyond the child's 26th birthday (White House 2010).

[9] As discussed in the empirical section, we also test in Table 4 the robustness of our main findings to other reasonable permutations of ages in the treatment and control groups and find that these choices do not change our conclusions.

The specification includes state fixed effects ($\delta_s$), year fixed effects ($\mu_t$), age fixed-effects ($\pi_g$), time-varying state-level controls ($x'_{st}$), as well as individual-level controls ($x'_{igst}$). We do not include $Treat_{ig}$ and $Post_t$ separately in the model because $Treat_{ig}$ is perfectly collinear with the age fixed effects $\pi_g$ while $Post_t$ is perfectly collinear with the year fixed effects $\mu_t$. The vector of individual controls $x'_{igst}$ includes race, ethnicity, education (Bachelor's degree or higher), and language spoken. The vector of time-varying state controls $x'_{st}$ includes income per capita, unemployment rate, state population size, racial, ethnic and age composition, percentage of state population with positive income from any state or local public assistance or welfare program, and cohabitation rate among different-sex couples. All specifications also account for LGBTQ policy changes: constitutional and statutory bans on same-sex marriage, same-sex marriage legalization, same-sex domestic partnership legalization, same-sex civil union legalization, LGBTQ non-discrimination laws, and LGBTQ hate crime laws. We also include controls for other relevant state policies: ACA Medicaid expansions and Medicaid private options.

This specification is estimated using only the sample of (married and unmarried) same-sex or different-sex couples. We estimate each specification separately for men and women. Standard errors are clustered at the level of the treatment: age (Abadie et al. 2017; Bertrand et al., 2004).[10] All specifications are weighted using the ACS person weights computed by the U.S. Census Bureau.

## 5. Results

Below, we present a collage of evidence on the effects of the ACA dependent coverage provision on health insurance coverage for individuals in same-sex couples. We begin by showing raw trends in health insurance outcomes, separately by gender and whether the individual is in a same-sex couple. We then turn to difference-in-differences regression results that compare changes in these outcomes for age-eligible (age 21-25) and slightly older (age 27-31) individuals in same-sex couples, and we do the same exercise for individuals in different-sex couples. We then present a range of robustness analyses – including event study regression estimates – that confirm the increases in health insurance we document for men in same-sex couples are real. Finally, we present a range of analyses that shed light on the mechanisms underlying the effects on insurance.

---

[10] All reported estimates have been computed using Stata 15. Given the small number of clusters, Stata automatically corrects critical values and p-values using - instead of a standard normal distribution - a T-distribution with degrees of freedom equal to the number of clusters minus one (Cameron et al. 2008).

### 5.1 Descriptive statistics and trends

Table 1 presents descriptive statistics for married and cohabiting young adults in the ACS. It shows that the vast majority of cohabiting young adults have health insurance, while a lower share (but still a majority) have employer-sponsored insurance. The majority of the sample is white and employed.

Figure 1 presents raw trends in the likelihood of any health insurance coverage for young adult men in same-sex couples (upper left panel), young adult men in different-sex couples (upper right panel), young adult women in same-sex couples (lower left panel), and young adult women in different-sex couples (lower right panel), separately by whether the individual is in the treatment age group or the control age group. Several patterns are apparent. First, health insurance coverage rates for individuals in same-sex couples were substantially lower than the associated rates for individuals in different-sex couples, especially in the early part of the sample period. This supports prior research showing disparities in health insurance coverage by sexual orientation. Second, younger individuals in both same-sex and different-sex couples both had lower rates of health insurance coverage than their slightly older counterparts in the early part of the sample period. Third, these gaps fell substantially beginning around 2011, consistent with an important role of the ACA dependent coverage provision extending parental ESI access to young adults. Finally, although there are only two data points prior to the ACA dependent coverage provision, there are not obviously different pre-treatment trends across the treatment (21 to 25-year-old) and control (27 to 31-year-old) groups.

Figure 2 plots the same rates for employer-sponsored insurance, and the format of Figure 2 is identical to that of Figure 1. The patterns in Figure 2 are broadly similar to those observed in Figure 1, though there is much less consistent evidence of a sexual orientation-related difference in employer-sponsored insurance for the younger individuals than there was in the likelihood of any insurance in Figure 1.[11] Overall the patterns in Figures 1 and 2 support a visual role for the ACA dependent coverage provision at increasing health insurance coverage for young adults aged 21-25 years in same-sex and different-sex couples. Moreover, there is some visual support for the idea that the ACA dependent mandate helped close gaps in health insurance coverage between adults in same-sex couples and adults in different-sex couples. We formalize and test for these differences in a regression framework in the next section.

---

[11] The gap in the likelihood of having any health insurance during the pre-treatment period for 21-25 year old men in same-sex couples compared to men in different-sex couples is driven by a much higher likelihood of reporting Medicaid coverage for men in different-sex couples compared to men in same-sex couples.

## 5.2 Effects of the ACA Dependent Coverage Provision on same-sex couples

Table 2 presents our baseline estimates of the effects of the ACA dependent coverage provision on the likelihood of any insurance coverage (columns 1, 3, and 5) and employer-sponsored insurance coverage (columns 2, 4, and 6).[12] We present results for men in the top panel and for women in the bottom panel. We present difference-in-differences results for individuals in same-sex couples in columns 1 and 2, and for comparison purposes we present the associated difference-in-differences results for individuals in different-sex couples in columns 3 and 4. These difference-in-differences models include all the individual controls described above, as well as the state/time varying controls for state demographic and economic characteristics and state LGBTQ policy environments. In columns 5 and 6 of Table 2 we report estimates from a fully interacted triple difference model where we test whether the insurance changes experienced by same-sex couples in columns 1-2 were meaningfully different from those experienced by different-sex couples in columns 3-4 by showing the coefficient on the triple interaction among being in the treatment group (age 21-25), being observed after 2010, and being in a same-sex couple (in a model that also controls for all the two-way interactions). In each panel we also report the mean of the dependent variable for the treatment group (age 21-25) prior to the reform (2008-2009).

The results in the top panel of columns 1 and 2 of Table 2 confirm the trends highlighted in Figures 1 and 2: the ACA dependent coverage provision was associated with an 8 percentage point increase in the likelihood that young men in same-sex couples aged 21-25 years reported having any health insurance coverage compared to the associated change for men in same-sex couples who were slightly older (age 27-31), and this estimate is statistically significant at the one percent level. Relative to the mean of the dependent variable for age-eligible men in same-sex couples prior to the reform, this is approximately a 12.8 percent effect. The results in the top panel of column 2 of Table 2 indicate that there was an even larger estimated average increase (11.1 percentage points) in the likelihood of employer-sponsored insurance for age-eligible men in same-sex couples, and this estimate is also statistically significant at the five percent level. Relative to the average of employer-sponsored insurance for age-eligible men in same-sex couples prior to the ACA dependent coverage provision, this is an even larger relative effect (23.4 percent).

---

[12] Prior research has examined whether the ACA dependent mandate affected household structure and marital status outcomes (Abramowitz 2016). In results not reported but available upon request, we also tested whether the ACA dependent coverage provision affected the likelihood of being in a same-sex couple. It is plausible that age-eligible individuals in dating relationships would have previously formed a cohabiting partnership with their romantic partner in order to gain health insurance (if the partner had a job with generous insurance, for example). After the ACA dependent coverage provision, these individuals might choose to get insurance from their parents and delay cohabitation with their romantic partner. If so, this would induce composition bias and affect interpretation of our core difference-in-differences models. We estimated equation (1) where the outcome is an indicator for being in a same-sex unmarried/married partnership and the sample is individuals in same-sex unmarried/married partnerships and single household heads, separately for men and for women. We found no statistically significant relationship between the ACA dependent coverage provision and this outcome for men or women, suggesting that composition biases are unlikely in our setting.

Turning to the difference-in-differences results for women in same-sex couples in the bottom panel of Table 2, we find smaller point estimates that are not statistically significant, though they are both positive in sign, consistent with the idea that the ACA dependent coverage provision increased insurance coverage for women in same-sex couples. The point estimate in the bottom panel of column 2 of Table 2, for example, indicates that the ACA dependent coverage mandate increased the likelihood that a woman aged 21-25 years in a same-sex couple had employer sponsored insurance by 3.5 percentage points, or 7.3 percent relative to the pre-reform mean for age-eligible women in same-sex couples. Thus, while we lack precision to identify statistically significant effects for women in same-sex couples, the evidence suggests a protective role for the ACA dependent coverage mandate for this group.

These estimates are broadly consistent with prior literature on the effects of the ACA dependent coverage mandate. Antwi et al. (2013) estimates that the dependent coverage provision increased the likelihood of any insurance coverage by three percentage points and the likelihood of having employer-sponsored dependent insurance by seven to ten percentage points using the Survey of Income and Program Participation. Barbaresco et al. (2015) find that the ACA dependent coverage provision increased the likelihood of any health insurance coverage by six percentage points using the Behavioral Risk Factor Surveillance System. Sommers et al. (2013) use data from the National Health Interview Survey and find increases in insurance coverage of about five percentage points associated with the ACA dependent coverage provision. Thus, our core estimates for men in same-sex couples are similar in magnitude to existing estimates from the prior literature.

### 5.3 Event study

We present standard event study estimates in Figures 3 and 4 for any health insurance and employer sponsored insurance, respectively, for individuals in same-sex couples (men in the top panel and women in the bottom panel). In these models we replace the indicator for "after 2010" with a series of event time indicators, interacting each ACS year with an indicator for treatment group observations (i.e., individuals age 21-25). Formally, we estimate the following model:

$$y_{igst} = \alpha + \sum_{k=2008}^{2018} \beta_k \left(Treat_{ig} * Year_k\right) + \delta_s + \mu_t + \pi_g + x'_{st}\gamma_1 + x'_{igst}\gamma_2 + \varepsilon_{igst}$$

All regressors are defined as in Section 4. As usual in the literature, we have normalized the first lead operator (the interaction with $Year_{2009}$) to zero. In line with the main specifications in Table 2, we have continued to exclude observations from 2010 in our analysis.

There is no evidence of differential pre-trends among respondents age 21-25 relative to those age 27-31 in any of the figures, thus supporting the parallel trend assumption in our difference-in-differences strategy. Moreover, the effect of the ACA dependent coverage provision appears nearly immediately (by 2011) for men in same-sex couples for both any insurance and employer-sponsored insurance. For men in same-sex couples, several individual event-time interactions are individually statistically significant.

For women in same-sex couples in Figures 3 and 4 we similarly observe no evidence of differential pre-trends, and there is also visual evidence of an increase in both any insurance coverage and employer-sponsored insurance in the years after 2010. Some of the individual post-ACA interaction terms are themselves individually significant.

## 5.4 Effect on different-sex couples and triple difference estimates

Columns 3 and 4 of Table 2 present the associated results on individuals in different-sex couples to benchmark the relative magnitudes of the effects of the ACA dependent coverage provision. Notably, in line with the previous literature and the trends in Figures 1-2, the pre-reform means for any insurance in column 3 for individuals in different-sex couples are substantially higher than the associated means for individuals in same-sex couples in column 1. For men in different-sex couples we estimate an increase in any insurance coverage of 1.2 percentage points, with a 3.8 percentage point increase in employer-sponsored insurance. Relative to the pre-reform means, these estimates correspond to 1.7 and 7.8 percent relative effects, respectively. For women the corresponding estimates are 2.6 and 2.8 percentage point increases (3.5 and 5.5 percent relative effects), respectively. All the difference-in-differences estimates for individuals in different-sex couples in columns 3 and 4 are statistically significant at the one percent level.[13]

Although the magnitude of the insurance increases for men in same-sex couples in the top panel of columns 1 and 2 of Table 2 is much larger than the associated increases for men in different-sex couples in the top panel of columns 3 and 4 of Table 2, in columns 5 and 6 we present triple difference models to explicitly test whether the increase in health insurance coverage for individuals in same-sex couples associated with the ACA dependent coverage provision was statistically different than the associated change for individuals in different-sex couples. Each entry in columns 5 and 6 is the coefficient on a triple interaction term among the indicators for being the treatment age group (21-25 years), being observed after 2010, and being in a same-sex couple. Formally, we estimate the following model:

$$y_{igstk} = \alpha + \beta\big(Treat_{ig} * Post_t * SameSex_{ik}\big) + \mu_{gt} + \pi_{kt} + \rho_{gk} + \delta_s + x'_{st}\gamma_1 + x'_{igstk}\gamma_2 + \varepsilon_{igstk}$$

---

[13] As an alternative way to benchmark the effect size for heterosexual individuals, we examined a sample of all household heads who reported being single. Since we know from other data that the share of individuals who identify as heterosexual is around 95 percent in most credible population-based datasets (Gates 2011), the vast majority of single household heads are likely to be heterosexual. We present those estimates in Appendix Table B1, which indicate that the ACA dependent coverage provision increased the likelihood of any health insurance coverage among single household heads by about 3.6 percentage points for both men and women, with larger increases in employer sponsored insurance (5.9 and 5.3 percentage points for men and women, respectively). These estimates are slightly larger than the associated difference-in-differences estimates for individuals in different-sex couples in columns 3 and 4 of Table 2, but the estimates for men are notably smaller than the difference-in-differences estimates for men in same-sex couples in the top panel of columns 1 and 2 of Table 2.

where $y_{igstk}$ is whether individual *i* in age group *g* living in state *s* at time *t* had any health insurance coverage (or employer-sponsored insurance). The subscript k indicates whether an individual is in a same-sex or different-sex couple. The coefficient of interest is $\beta$. $Treat_{ig}$ and $Post_t$ are defined as in Section 4 and interacted with the same-sex couple indicator $SameSex_{ik}$. The specification includes age-specific time effects that are common across couples ($\mu_{gt}$), time-varying effects specific to same-sex couples ($\pi_{kt}$), age-specific effects among same-sex couples ($\rho_{gk}$), state fixed effects ($\delta_s$), state controls ($x'_{st}$), and individual controls ($x'_{igstk}$). We do not include the double-interactions between $Treat_{ig}$, $Post_t$, and $SameSex_{ik}$ since they are perfectly collinear with the fixed effects $\mu_{gt}$, $\pi_{kt}$, and $\rho_{gk}$.

We emphasize here that these triple difference estimates are presented for descriptive purposes only. That is, we are not arguing that additionally differencing out the effect for individuals in different-sex couples allows us to more accurately estimate the true causal effect of the ACA dependent coverage provision on individuals in same-sex couples, and we recognize that pathways into and out of relationships for sexual minorities and heterosexual individuals may differ for any number of reasons, including possibly due to the roles of social and policy context. Instead, we present these triple difference estimates as another interesting benchmark for understanding the strength and magnitude of the ACA dependent mandate effects on individuals in same-sex couples.

The findings in the top panel of columns 5 and 6 of Table 2 indicate that the increases in the likelihood of any insurance coverage for men in same-sex couples associated with the ACA dependent coverage provision were, in fact, significantly larger than the associated increases for men in different-sex couples. For any health insurance, for example, we estimate that age-eligible men in same-sex couples experienced an increase of 6.5 percentage points greater than what was experienced by age-eligible men in different-sex couples coincident with the ACA dependent coverage provision. We estimate a similarly sized 6.1 percentage point triple interaction for employer-sponsored insurance in the top panel of column 6, but it is not statistically significant. For women (presented in the bottom panel of Table 2) we find much smaller triple difference estimates, and neither is statistically significant.

## 5.5 Extensions and robustness checks

In Table 3 we present the associated results for outcomes reflecting the other sources of health insurance. We present results from the specification in columns 1-4 of Table 2 with the main effects, individual controls, and state/time varying controls, and we present the coefficient on the interaction term between the indicators for age 21-25 years and after 2010. As in Table 2, we present results for men in same-sex couples in the top panel and for women in same-sex couples in the bottom panel. We reprint the estimates for having any health insurance and for having employer-sponsored insurance in columns 1 and 2, respectively, and we present results for direct/privately purchased insurance in column 3, for Tricare in column 4, for Medicare in column 5, for Medicaid in column 6, for Veterans Affairs (VA) coverage in column 7, and for Indian Health Service coverage in column 8.

The results in the top panel of Table 3 suggest that the discrepancy between the larger increase in employer-sponsored insurance and the increase in the likelihood of any health insurance for men in same-sex couples associated with the ACA dependent coverage mandate is due in part to a large reduction in the uninsurance rate and a sizable reduction in the likelihood of reporting Medicaid coverage (though the Medicaid estimate is not statistically significant). These results suggest that the ACA's dependent coverage provisions were effective at lowering the uninsurance rate for young men in same-sex couples. Meanwhile, the 'reverse crowd out' phenomenon (i.e., increases in private health insurance that are coincidentally associated with decreases in public health insurance) has been documented in previous research on state-level dependent coverage provisions (Levine et al. 2011). Coefficient estimates on the other sources of insurance are very small and not statistically significant. For women in same-sex couples in the bottom panel of Table 4, we continue to find no evidence of statistically significant changes in health insurance coverage associated with the ACA dependent coverage provision except for a marginally significant reduction in Tricare coverage, in line with the decline in military participation among young adults following the ACA reform documented by Chatterji et al. (2019).

In Table 4 we present robustness checks where we vary the ACS years used in the analysis (columns 1-3) and the age-based definitions of treatment and control groups (columns 4-6) for the outcome of any health insurance. We restrict attention to individuals in same-sex couples, and we present results for men in the top panel and for women in the bottom panel. Each column header describes the sample restriction that we impose. The patterns in Table 4 confirm that the finding of increased health insurance for men in same-sex couples associated with the ACA dependent coverage provision is highly robust to reasonable alternative choices about which years of the ACS to include and which ages should constitute treatment and control groups. In every case we find that the ACA dependent mandate is associated with large and statistically significant increases in the likelihood of having health insurance for men in same-sex couples.[14] This pattern is reassuring given that some prior research on the ACA dependent coverage provision has documented sensitivity of findings on health insurance coverage to these alternative choices (Slusky 2017). For women, we continue to find suggestive—but not statistically significant—evidence of increases in health insurance coverage associated with the ACA dependent coverage provision, except for the full period 2008-2018 which does return a marginally significant increase in insurance coverage of 6.3 percentage points (or 9.6 percent of the pre-reform mean for the treatment group).[15]

---

[14] The larger estimates when including respondents in later years could be due to the fact that some insurance plans ('grandfathered employer plans') were allowed to refuse coverage to age-qualified dependent children whose own employers offered them health insurance until 2014 (Antwi et al. 2013).

[15] As a placebo test, we have also compared changes in insurance coverage between individuals age 27-31 and those age 32-36 before and after 2010. The estimated difference-in-difference coefficient in Appendix Table B2 is small and statistically insignificant for both men and women in same-sex couples, when looking at either the probability

In Table 5, we present a series of additional robustness checks and extensions for our main results for men in same-sex households. We vary the format of Table 5 slightly in that we focus only on men in same-sex households – the group for whom we find the most consistent evidence of protective effects of the ACA dependent coverage mandate – and present results for any insurance in the top panel and for employer-sponsored insurance in the bottom panel. In Column 1 of Table 5 we show results from a model where instead of controlling for time-varying state characteristics we include a full set of state-by-year fixed effects. In this flexible model we continue to find that the ACA dependent coverage provision was associated with even larger and statistically significant increases in health insurance coverage and employer sponsored insurance for men in same-sex couples.

In column 2 of Table 5, we show results from a sample that excludes the handful of states that had legal access to same-sex marriage before 2010, and in column 3 of Table 5 we show results from a sample that excludes states that had legal access to same-sex marriage at any time during our 2008-2012 sample period. Neither sample restriction meaningfully changes the core finding, which is important and suggestive that young men in same-sex couples could be enrolled in a parent's ESI plan rather than a spouse's ESI plan. This robustness is not particularly surprising since the research design hinges on over-time comparisons across slightly younger and slightly older young adults, and thus it is difficult to think about confounding factors that differentially affected these two groups.[16]

**5.6 Suggestive evidence on the underlying mechanisms**

Having documented a robust increase in the likelihood of having any health insurance coverage and employer-sponsored insurance for men in same-sex couples associated with the ACA dependent coverage provision, that in some cases is significantly larger than the same effect enjoyed by men in different-sex couples, we turn the focus of our analysis in Table 6 to several tests that help us further understand mechanisms and plausibility. The format of Table 6 follows that of Table 5 in that we concentrate on men in same-sex couples and report results for any health

---

of having any insurance coverage or employer-sponsored insurance, thus supporting our identification strategy and the claim that the estimated increase in health insurance coverage among respondent age 21-25 is causal and not resulting from a spurious relationship.

[16] The Appendix reports the results of several other robustness tests we performed on the main results reported in columns 1 and 2 of Table 2. Appendix Table B3 shows that our main results are robust to clustering standard errors at the state level (as in Antwi et al., 2013), to estimating heteroscedasticity-robust standard errors, to estimating p-values using the wild cluster bootstrap procedure (MacKinnon and Webb 2018; Roodman et al. 2019), to estimating p-values using the effective number of clusters (Carter et al. 2017; Lee and Steigerwald 2018), to estimating models without the ACS person weights, and to estimating models using the ACS replication weights. Appendix Table B4 shows that our main results are also robust to excluding same-sex spouses from the 2012 estimation sample and only examining individuals in same-sex unmarried partnerships to address concerns about misclassification errors being more common among married couples (O'Connell and Feliz, 2011), to including 2010 ACS data and counting that year as treated by the ACA dependent coverage provision, to including 2010 ACS data and coding that year as untreated, to including 26-year-old respondents as part of the control group, and to restricting attention to individuals age 23-25 versus 27-29 as suggested by Slusky (2017). Appendix Table B5 shows that our main results for men are robust to excluding each individual state one at a time.

insurance in the top panel and for employer-sponsored insurance in the bottom panel. In columns 1 and 2 of Table 6, we show results separate for individuals whose state of residence at the time of the interview was equal or not equal to their reported state of birth, respectively. Although out-of-state migration is correlated with many important unobservable characteristics (including, presumably, sexual orientation), we note that pre-reform means of the outcome variables are quite similar across these two groups and certainly smaller than the differences between individuals in same-sex couples and individuals in different-sex couples in Table 2. We hypothesize that individuals who had not migrated from their state of birth were more likely to be physically proximate to their parents, thus reducing the cost of accessing dependent coverage. Non-migration since birth may also signal stronger family relationships. Indeed, we observe much larger effects for non-migrants than for migrants.

In columns 3 and 4 of Table 6 we present results separately for individuals who are the household head (i.e., the primary reference person in whose name the property is owned or rented) versus the partner or spouse of the household head, respectively. A stark pattern emerges: all of the effect of the ACA dependent coverage provision accrues to partners of household heads, with no effect on the household heads themselves. There are several possible explanations for these results. First, it could be that the household heads had employer-sponsored insurance that did not cover family members. Second, it could be that the household heads had employer-sponsored insurance that covered some family members but did not cover same-sex partners. While large firms over this time period were increasingly offering health insurance benefits to same-sex unmarried partners, coverage was far from universal. In fact, Dawson et al. (2016) found that in 2016 only 43% of firms offering spousal benefits had extended such coverage to same-sex spouses. Third, it could be that the household heads did not want to effectively out themselves to their employers as being sexual minorities, which they would have had to do in order to claim same-sex partners as dependents for health insurance purposes. Without additional data, we cannot directly test which of these channels was driving this pattern.[17]

In Table 7 we further explore mechanisms by examining other possible margins of adjustment. Specifically, we examine employment and student status. We hypothesize that the increased access to parental health insurance coverage via the ACA dependent coverage mandate allowed individuals to reduce employment (if they were working primarily to obtain health insurance) and/or increase schooling. It is worth remembering that some prior dependent coverage mandates at the state level imposed requirements such as enrolling in school and/or being unmarried (in addition to being below a certain age threshold). We report these results in Table 7, with effects for men in same-sex couples in the top panel and for women in same-sex couples in the bottom panel. Each column shows the results from the standard difference-in-differences specification for various indicator variables: being employed (in the prior week) in column 1, being unemployed in column 2, being in the labor force (either employed or unemployed) in column 3, working at least 30 hours per week in column 4, working at least 40 hours per week in column 5, and being a student within the past three months in column 6.

---

[17] In Appendix Tables B6 and B7 we investigated heterogeneity in the results for men in same-sex couples with respect to education and race, respectively. Table B6 shows that the increases in insurance coverage experienced by men in same-sex couples associated with the ACA dependent coverage mandate were observed primarily for individuals without a Bachelor's degree. Table B7 shows that the increases in insurance coverage are statistically significant only for white men in same-sex couples, though the point estimates for the other race groups are in some cases large and positive even when they are not statistically significant.

The patterns in Table 7 reveal that the ACA dependent coverage provision had little effect on employment or labor force attachment or school enrollment for men in same-sex couples in the top panel. All estimates are small and statistically insignificant. For women in same-sex couples in the bottom panel, in contrast, we estimate that the ACA dependent coverage provision was associated with statistically significant reductions in the likelihood of working at least 40 hours per week (column 4) and with a statistically significant reduction in total work hours of about 4.5 hours (column 5). This pattern is consistent with the lack of an overall change in employer-sponsored insurance for women in same-sex couples and suggests that women in same-sex couples may have traded own employer-sponsored insurance for parental coverage in response to the ACA dependent coverage mandate. The reductions in full-time work are accounted for by women in same-sex couples having increased risk of being unemployed (column 2), exiting the labor force (column 3), and being a student (column 7), though not all of these estimates are statistically significant.

## 6. Discussion and conclusion

A large body of prior research documents that the dependent coverage provision of the Affordable Care Act was associated with meaningful increases in health insurance coverage for young adults after it took effect in 2010. We provide the first examination of whether young adults in same-sex couples – the vast majority of whom are likely to be gay, bisexual, queer, and lesbian – also benefitted from this reform. We hypothesized that a higher likelihood of strained relationships with parents might mean that sexual minorities in same-sex couples would have lower opportunity to benefit from the dependent coverage provision. Perhaps surprisingly, then, we found that young adults in same-sex couples who were age-eligible for the ACA dependent mandate experienced significant increases in health insurance coverage after 2010 compared to the associated change for their slightly older counterparts who were not eligible to gain parental coverage. This increase was driven by large improvements in the likelihood of having employer-sponsored insurance. The effects we identify were consistently observed for young men in same-sex couples, with smaller effects that were not always statistically significant for young women in same-sex couples.

How large are the effects we identify? Consider that from 2008-2018 the share of young men in same-sex couples aged 21-25 years who reported employer-sponsored insurance increased by about 24 percentage points (upper left panel of Figure 1). When measured over the full sample period, we estimate that the ACA dependent mandate significantly increased the likelihood of employer sponsored insurance by 11.8 percentage points (top panel of column 3 of Table 4). Thus, we estimate that the ACA dependent coverage provision can account for about half of the increase in overall health insurance coverage for young men in same-sex couples over this time period.

We also found that the increase in health insurance we identify for men in same-sex couples is significantly larger than the associated increase for men in different-sex couples. Why might this be the case? There are several possibilities, though we do not have data to adjudicate among them. First, as noted above, men in same-sex couples who were not the household head may have had greater need for parental health insurance coverage due to lack of access to the employer-sponsored insurance of their partners/spouses. Even if they did have partners/spouses with employer sponsored insurance coverage that would have extended to same-sex partners, they may have feared employer-based discrimination or other reprisals by taking it up. Second, men in same-sex couples may have had higher demand for health insurance because of the differential burden of some health conditions within the sexual minority male community, including HIV and poor mental health. These factors may have contributed to the larger effects of the ACA dependent coverage mandate on insurance coverage for men in same-sex couples compared with men in different-sex couples.

Regarding women in same-sex couples, we found weaker evidence of increases in health insurance associated with the ACA dependent coverage provision, though several patterns point to improvements that were smaller in scale than those we identify for men in same-sex couples. First, the point estimates from our main specification in Table 2 – though not statistically significant – were sizable as a share of the pre-reform mean, especially for employer-sponsored insurance (an estimated 7.3 percent relative increase for 21-25 year old women in same-sex couples after 2010 compared to the associated change for 27-31 year old women in same-sex couples). Second, Table 4 showed that lengthening the time period under study returned successively larger estimates of the protective effect of the ACA dependent coverage provision on the likelihood of any health insurance coverage for women in same-sex couples, such that the estimates attained marginal statistical significance when we considered the longest period (2008-2018). Third, further robustness analyses of the results for women shown in Appendix Table B8 demonstrate that the increases in insurance coverage for women in same-sex couples associated with the ACA dependent coverage mandate were much larger for women who did not migrate from their state of birth than for women who did migrate from their state of birth, similar to the patterns we observed for men in same-sex couples in Table 6. Finally, Appendix Table B8 also confirms that the ACA dependent coverage mandate was associated with statistically significant increases in the likelihood of having employer-sponsored insurance for women in same-sex couples who were the partners of the household head (but not for women in same-sex couples who were themselves the household heads). This pattern exactly matches the pattern for men in same-sex couples in Table 6. Thus, taken together, we conclude that there are several patterns suggesting that the ACA dependent coverage provision also increased insurance coverage for women in same-sex couples, though these effects are consistently smaller than those observed for men in same-sex couples. These findings are similar to those in other studies in the literature: both Antwi et al. (2013) and Barbaresco et al. (2015) also found larger effects for men than for women associated with the ACA dependent coverage provision, even if they did not specifically examine individuals in same-sex couples.[18]

Our study is subject to several limitations, many of them owing to challenges in identifying sexual minorities in the ACS. First, because the ACS does not include direct questions about sexual orientation at the individual level, we cannot identify effects of the ACA dependent coverage provision on health insurance coverage of single sexual minorities. It could be that being in a same-sex couple signals some positive relationship with family members (i.e., perhaps the sexual minorities who have difficult relationships with parents are less likely to be coupled). Related to this, despite documented disparities in health for transgender individuals (Lagos 2018), we have no information on gender identity, and so we cannot address the effects of the ACA on transgender populations, who may also have strained relationships with their parents and unique healthcare needs. A related limitation of relying on relationships to the ACS household head to identify same-sex couples is that if an unmarried same-sex couple moved in with one of the couple's parents, it would be very unlikely that we could identify them as a same-sex couple. In that situation the household head would likely be the parent, not the member of the same-sex couple, and one member of the couple would be identified as son or daughter but the other member of the couple would most likely be identified as 'other nonrelative'. That is, if the same-sex couple does not involve the householder, there is no way to identify in the ACS that those two individuals in the same-sex couple are in a romantic relationship.[19]

Second, although the ACS permits us to identify different types of health insurance, for employer-sponsored insurance, we do not know the name of the person in whose name the employer policy is written (i.e., the policyholder). Because of this, we can speculate that unmarried partner men age 21-25 in same-sex couples are gaining health insurance from their own parent, but we cannot directly confirm this. Of course, we can think of no other confounding policy or other variable that would differentially affect individuals aged 21-25 compared to those aged 27-31 coincident with the 2010 ACA dependent mandate, and so we are leaning heavily on the difference-in-differences design in this case. Third, the ACS lacks information on access to care, health services utilization, and health outcomes, and so we can only examine effects on health insurance coverage. We leave examination of these other health outcomes to future research.

---

[18] From a statistical point of view, it is worth emphasizing that the confidence intervals for the estimated impacts of the ACA dependent coverage mandate on women in same-sex couples are often very large and overlapping with those for men in same-sex couples. Nevertheless, there are many substantive reasons why the estimated effects could be larger for sexual minority men in same-sex couples than for sexual minority women in same-sex couples. For example, most of the labor economics literature shows that gay men suffer a wage penalty compared to comparably skilled heterosexual men, while lesbians earn a wage premium compared to comparably skilled heterosexual women (Klawitter 2015; Neumark 2018), which is consistent with the idea that labor market discrimination against gay men is stronger than against lesbian women. This would be consistent with a greater need among men in same-sex couples for parental insurance coverage than among women in same-sex couples.

[19] Note moreover that this problem is more severe for sexual minorities than for heterosexuals, since if a different-sex couple chose to get married and move in with one of their parents, the different-sex spouse would be identified as son-in-law or daughter-in-law of the household head.

Despite these limitations, our findings confirm the broad effects of expanded dependent coverage and suggest that eliminating the federal dependent mandate could reduce health insurance coverage among young adult sexual minorities in same-sex couples. In so doing, our study also provides one of the literature's first quasi-experimental examinations of how population-targeted (i.e., not LGBTQ-specific) health policies affected sexual minorities, including whether it had differential effects relative to heterosexual populations. Social science and public health literatures have made important advances in documenting heterogeneous treatment effects by age, gender, race/ethnicity, and education across a range of important health and social policies. Our results highlight the importance of adding sexual orientation to that standard list of demographic characteristics in order to monitor and achieve health equity for LGBTQ people in the United States.

**Figure 1: Trends in health insurance rates. Individuals in same-sex couples (SSC) and different-sex couples (DSC).**

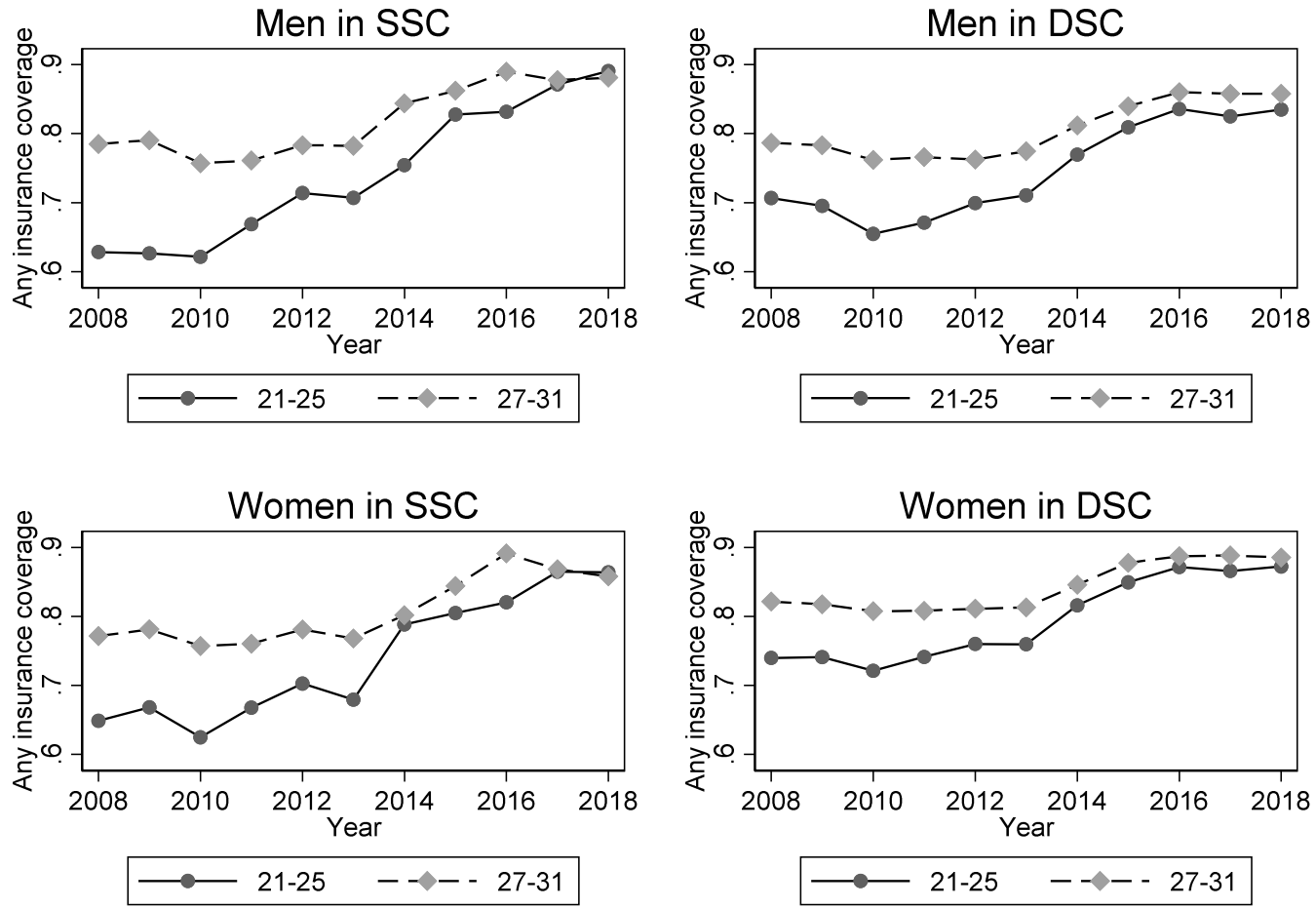

Notes: The dependent variable is whether the respondent had any health insurance coverage. "SSC" indicates same-sex couples. "DSC" indicates different-sex couples. Weighted summary statistics using person weights. Source: ACS 2008-2018.

**Figure 2: Trends in employer health insurance rates. Individuals in same-sex couples (SSC) and different-sex couples (DSC).**

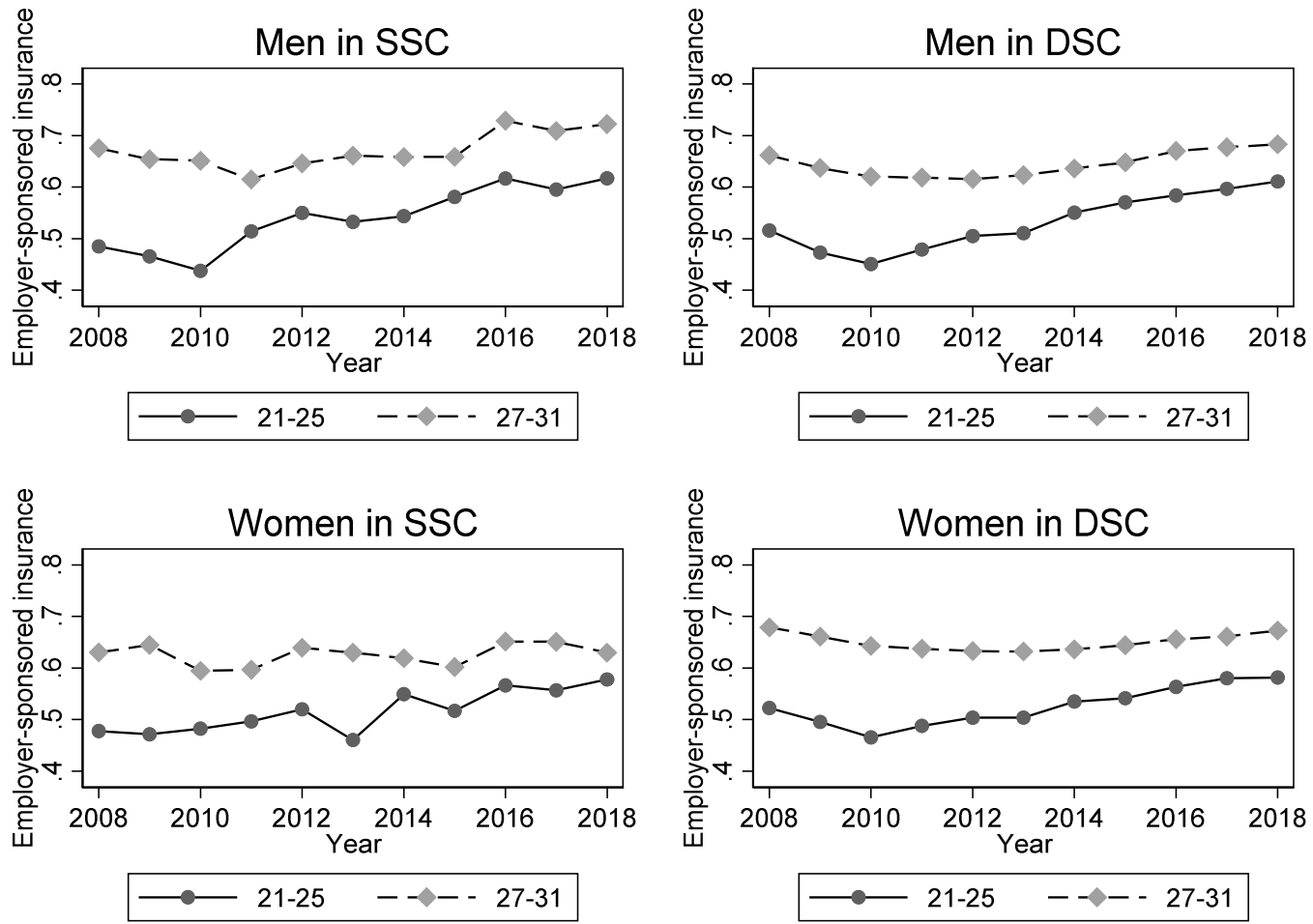

Notes: The dependent variable is whether the respondent had health insurance through an employer. "SSC" indicates same-sex couples. "DSC" indicates different-sex couples. Weighted summary statistics using person weights. Source: ACS 2008-2018.

**Figure 3: Event study estimates of the effect of ACA on any health insurance among individuals in same-sex couples.**

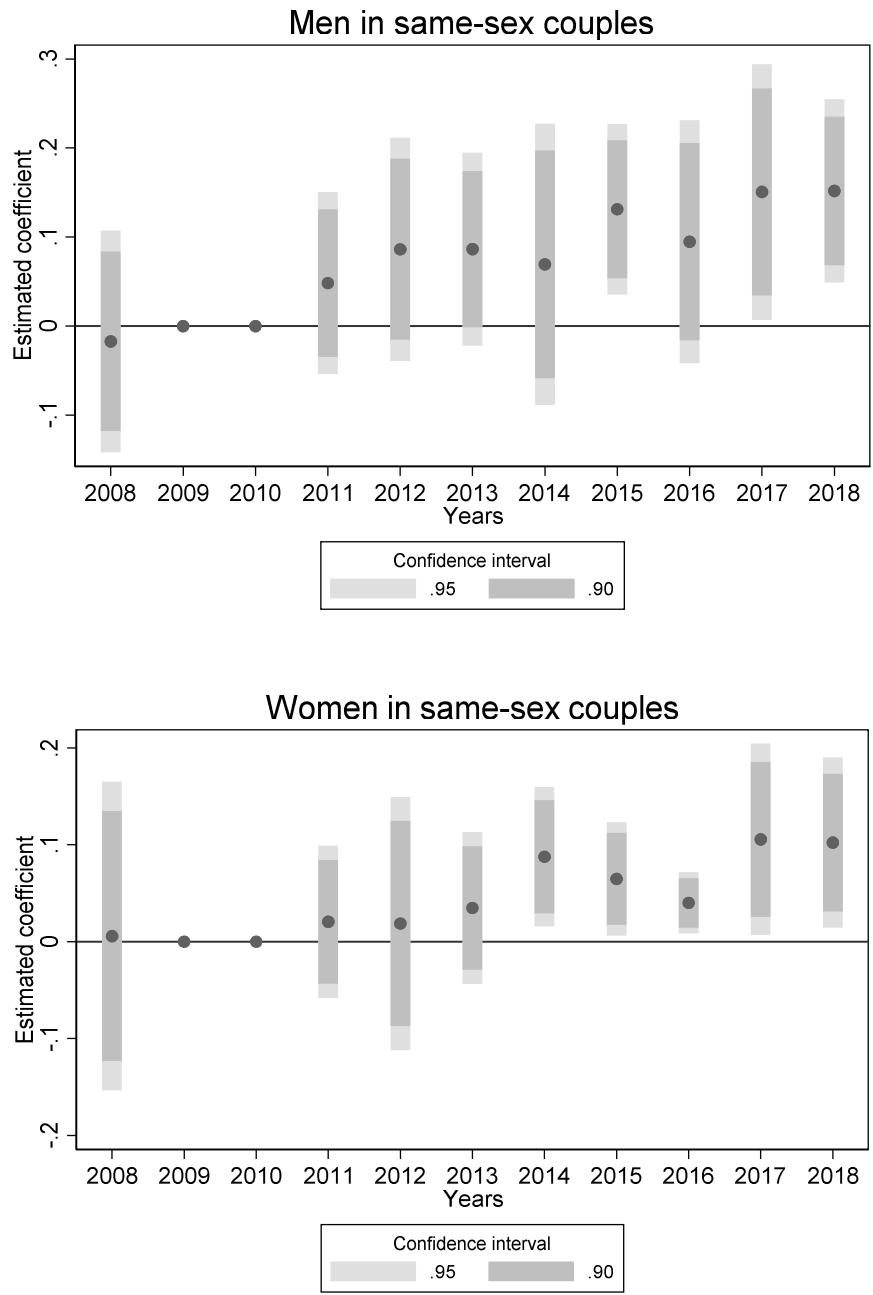

Notes: The dependent variable is whether the respondent had any health insurance coverage. Sample includes respondents in either married or unmarried same-sex couples. Individuals age 21-25 are compared to those age 27-31. Same fixed effects, individual and state controls as Table 2. Shaded bars represent the 90 and 95 percent confidence intervals. Weighted regressions using person weights. Source: ACS 2008-2018 (excluding 2010).

**Figure 4: Event study estimates of the effect of ACA on employer-sponsored insurance among individuals in same-sex couples.**

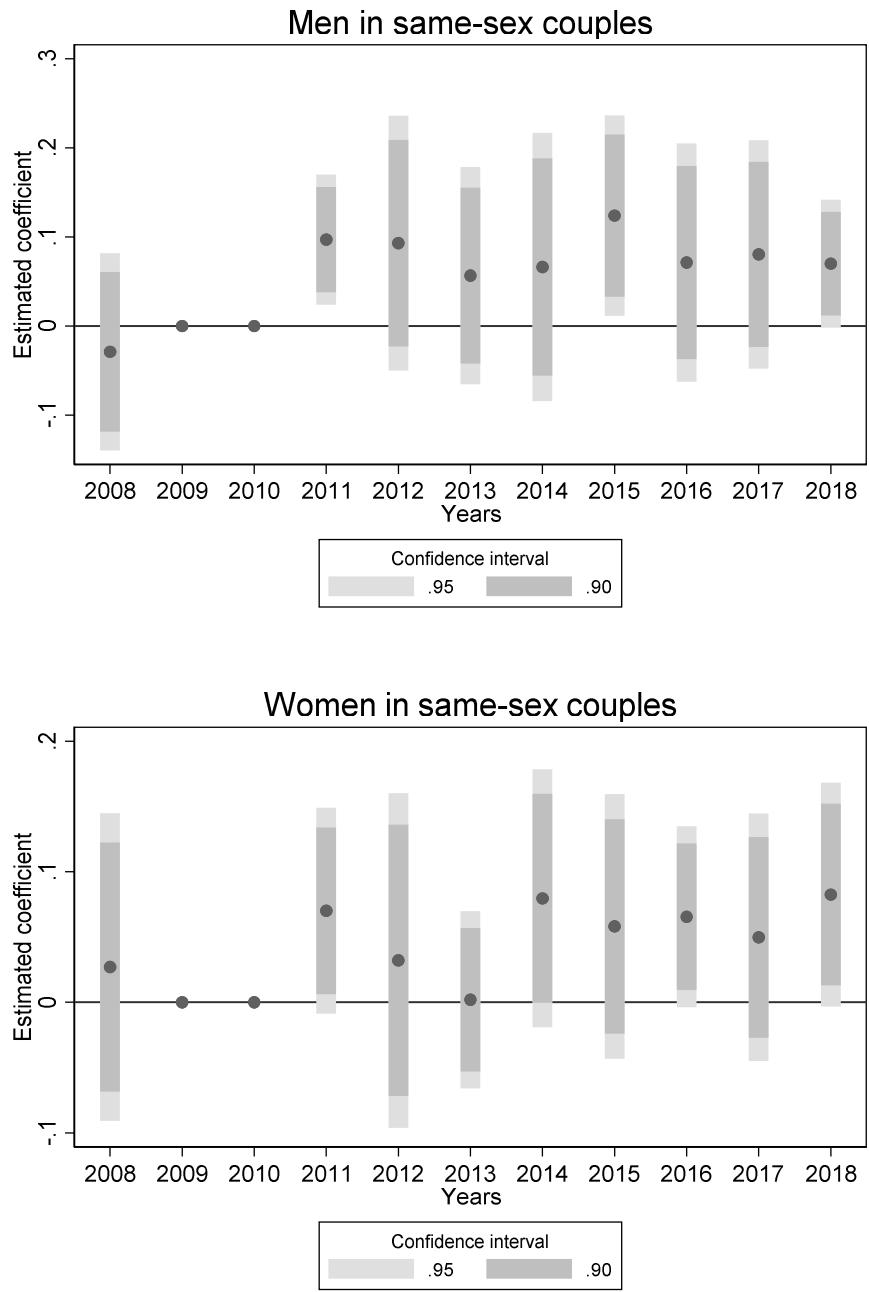

Notes: The dependent variable is whether the respondent had health insurance through an employer. Sample includes respondents in either married or unmarried same-sex couples. Individuals age 21-25 are compared to those age 27-31. Same fixed effects, individual and state controls as Table 2. Shaded bars represent the 90 and 95 percent confidence intervals. Weighted regressions using person weights. Source: ACS 2008-2018 (excluding 2010).

**Table 1: Summary statistics for individuals in same-sex and different-sex couples**

|  | Individuals in same-sex couples | | Individuals in different-sex couples | |
|---|---|---|---|---|
|  | Male | Female | Male | Female |
| *Main dependent variables:* | | | | |
| Has any health insurance coverage | 0.739 | 0.736 | 0.752 | 0.792 |
| Has employer sponsored insurance | 0.598 | 0.577 | 0.595 | 0.603 |
| | | | | |
| *Individual controls:* | | | | |
| White | 0.782 | 0.755 | 0.784 | 0.788 |
| Black | 0.089 | 0.122 | 0.086 | 0.068 |
| Asian | 0.038 | 0.027 | 0.036 | 0.055 |
| Other races | 0.091 | 0.096 | 0.093 | 0.088 |
| Hispanic | 0.182 | 0.150 | 0.197 | 0.184 |
| College education | 0.368 | 0.342 | 0.261 | 0.333 |
| Does not speak English | 0.013 | 0.004 | 0.013 | 0.021 |
| | | | | |
| *Other key characteristics:* | | | | |
| Employed (vs. Unemployed/NILF) | 0.822 | 0.814 | 0.888 | 0.681 |
| Unemployed (vs. Employed/NILF) | 0.065 | 0.084 | 0.063 | 0.056 |
| Work 30h/week or more | 0.802 | 0.779 | 0.901 | 0.630 |
| Work 40h/week or more | 0.664 | 0.622 | 0.812 | 0.481 |
| Student | 0.188 | 0.219 | 0.118 | 0.159 |
| Total personal income (pre-tax) | 33,911 | 26,270 | 38,064 | 22,297 |
| Observations | 2,781 | 3,614 | 235,954 | 304,318 |

Notes: Sample includes respondents in either married or unmarried same-sex and different-sex couples, aged 21-25 or 27-31 years. Weighted summary statistics using person weights. Source: ACS 2008-2012 (excluding 2010).

**Table 2: Effect of ACA dependent coverage mandate on health insurance for individuals in same-sex and different-sex couples.**

| | Individuals in same-sex couples | | Individuals in different-sex couples | | Individuals in same-sex and different-sex couples | |
|---|---|---|---|---|---|---|
| | Any insurance | Employer sponsored insurance | Any insurance | Employer sponsored insurance | Any insurance | Employer sponsored insurance |
| | (1) | (2) | (3) | (4) | (5) | (6) |
| *Men* | | | | | | |
| Age 21-25 * Post-2010 | 0.080*** | 0.111** | 0.012*** | 0.038*** | -- | -- |
| | (0.023) | (0.040) | (0.003) | (0.007) | | |
| Age 21-25 * Post-2010 * Same-sex | -- | -- | -- | -- | 0.065** | 0.061 |
| | | | | | (0.027) | (0.042) |
| N | 2,781 | 2,781 | 235,954 | 235,954 | 238,735 | 238,735 |
| Mean of DV for 21-25 pre-2010 | 0.627 | 0.474 | 0.701 | 0.495 | 0.627 | 0.474 |
| Adjusted R-squared | 0.176 | 0.161 | 0.134 | 0.127 | 0.134 | 0.127 |
| | | | | | | |
| *Women* | | | | | | |
| Age 21-25 * Post-2010 | 0.015 | 0.035 | 0.026*** | 0.028*** | -- | -- |
| | (0.037) | (0.042) | (0.006) | (0.005) | | |
| Age 21-25 * Post-2010 * Same-sex | -- | -- | -- | -- | -0.018 | -0.001 |
| | | | | | (0.038) | (0.041) |
| N | 3,614 | 3,614 | 304,318 | 304,318 | 307,932 | 307,932 |
| Mean of DV for 21-25 pre-2010 | 0.659 | 0.475 | 0.740 | 0.509 | 0.659 | 0.475 |
| Adjusted R-squared | 0.121 | 0.120 | 0.136 | 0.156 | 0.136 | 0.156 |
| | | | | | | |
| *Controls for:* | | | | | | |
| Age and year FE | X | X | X | X | | |
| State FE | X | X | X | X | X | X |
| Individual controls | X | X | X | X | X | X |
| State time-varying policies | X | X | X | X | X | X |
| Age-by-year, state-by-year, & age-by-state FE | | | | | X | X |

Notes: The dependent variable in columns 1,3, and 5 is whether the respondent had any health insurance coverage. The dependent variable in columns 2, 4, and 6 is whether the respondent had health insurance through an employer. Sample includes respondents in either married or unmarried different-sex or same-sex couples. Individuals age 21-25 are compared to those age 27-31. The mean of the dependent variable only refers to individuals age 21-25 interviewed in 2008 or 2009 (and only in same-sex couples in columns 5-6). Individual controls: ethnicity, race, language, education. State controls: income per capita, unemployment rate, population, racial and age composition, percentage of state population with positive welfare income, cohabitation rate among different-sex couples, constitutional and statutory bans on same-sex marriage, same-sex marriage legalization, same-sex domestic partnership legalization, same-sex civil union legalization, LGBTQ anti-discrimination laws, LGBTQ hate crime laws, Medicaid pre-expansion. Standard errors clustered at the age level in parentheses. Weighted regressions using person weights. Source: ACS 2008-2012 (excluding 2010). * $p < 0.10$, ** $p < 0.05$, *** $p < 0.01$

**Table 3: Effect of ACA on health insurance among same-sex couples, by type of coverage.**

|  | Any coverage | Employer | Direct purchase | TRICARE | Medicare | Medicaid | Veterans Affairs | Indian Health Service |
|---|---|---|---|---|---|---|---|---|
|  | (1) | (2) | (3) | (4) | (5) | (6) | (7) | (8) |
| *Men* | | | | | | | | |
| Age 21-25 * Post-2010 | 0.080*** | 0.111** | -0.010 | 0.005 | -0.004 | -0.013 | 0.001 | -0.003 |
|  | (0.023) | (0.040) | (0.026) | (0.013) | (0.006) | (0.015) | (0.009) | (0.005) |
| N | 2,781 | 2,781 | 2,781 | 2,781 | 2,781 | 2,781 | 2,781 | 2,781 |
| Mean of DV for 21-25 pre-2010 | 0.627 | 0.474 | 0.099 | 0.017 | 0.006 | 0.078 | 0.005 | 0.005 |
| Adjusted R-squared | 0.176 | 0.161 | 0.036 | 0.056 | -0.002 | 0.054 | 0.015 | 0.043 |
| *Women* | | | | | | | | |
| Age 21-25 * Post-2010 | 0.015 | 0.035 | -0.005 | -0.021* | -0.010 | -0.025 | 0.000 | 0.005 |
|  | (0.037) | (0.042) | (0.018) | (0.010) | (0.006) | (0.021) | (0.007) | (0.004) |
| N | 3,614 | 3,614 | 3,614 | 3,614 | 3,614 | 3,614 | 3,614 | 3,614 |
| Mean of DV for 21-25 pre-2010 | 0.659 | 0.475 | 0.098 | 0.025 | 0.012 | 0.129 | 0.002 | 0.000 |
| Adjusted R-squared | 0.121 | 0.120 | 0.029 | 0.052 | 0.005 | 0.086 | 0.047 | 0.056 |
| *Controls for:* | | | | | | | | |
| Age, State and year FE | X | X | X | X | X | X | X | X |
| State time-varying policies | X | X | X | X | X | X | X | X |
| Individual controls | X | X | X | X | X | X | X | X |

Notes: Sample includes respondents in either married or unmarried same-sex couples. Individuals age 21-25 are compared to those age 27-31. The mean of the dependent variable only refers to individuals age 21-25 interviewed in 2008 or 2009. Same fixed effects, individual and state controls as Table 2. Standard errors clustered at the age level in parentheses. Weighted regressions using person weights. Source: ACS 2008-2012 (excluding 2010). * $p<0.10$, ** $p<0.05$, *** $p<0.01$

**Table 4: Robustness of the effect of ACA on health insurance among same-sex couples with respect to sample years and treatment/control group ages.**

|  | Vary year range | | | Vary age range | | |
|---|---|---|---|---|---|---|
|  | 2008-2014 | 2008-2016 | 2008-2018 | 19-25 vs 27-33 | 20-25 vs 27-32 | 22-25 vs 27-30 |
|  | (1) | (2) | (3) | (4) | (5) | (6) |
| *Men* | | | | | | |
| Treated age group * Post-2010 | 0.090** | 0.100** | 0.118** | 0.093*** | 0.080*** | 0.059** |
|  | (0.033) | (0.034) | (0.037) | (0.022) | (0.021) | (0.019) |
| N | 4,611 | 6,950 | 9,712 | 3,653 | 3,254 | 2,257 |
| Mean of DV for treated age pre-2010 | 0.627 | 0.627 | 0.627 | 0.612 | 0.627 | 0.653 |
| Adjusted R-squared | 0.132 | 0.133 | 0.125 | 0.176 | 0.176 | 0.175 |
|  | | | | | | |
| *Women* | | | | | | |
| Treated age group * Post-2010 | 0.041 | 0.042 | 0.063* | 0.009 | 0.014 | 0.030 |
|  | (0.027) | (0.027) | (0.031) | (0.031) | (0.036) | (0.043) |
| N | 6,048 | 8,922 | 12,519 | 4,824 | 4,237 | 2,922 |
| Mean of DV for treated age pre-2010 | 0659 | 0.659 | 0.659 | 0.653 | 0.656 | 0.681 |
| Adjusted R-squared | 0.107 | 0.100 | 0.093 | 0.120 | 0.124 | 0.121 |
|  | | | | | | |
| *Controls for:* | | | | | | |
| Age, state and year FE | X | X | X | X | X | X |
| State time-varying policies | X | X | X | X | X | X |
| Individual controls | X | X | X | X | X | X |

Notes: The dependent variable is whether the respondent had any health insurance coverage. Sample includes respondents in either married or unmarried same-sex couples. Individuals age 21-25 are compared to those age 27-31 in columns 1-3. Column 4 compares individuals age 19-25 to those age 27-33. Column 5 compares individuals age 20-25 to those age 27-32. Column 6 compares individuals age 22-25 to those age 27-30. The mean of the dependent variable only refers to individuals in the treated age group interviewed in 2008 or 2009. Same fixed effects, individual and state controls as Table 2. Standard errors clustered at the age level in parentheses. Weighted regressions using person weights. Source: ACS 2008-2014 (Column 1), 2008-2016 (Column 2), 2008-2018 (Column 3); 2008-2012 (Columns 4-6). All specifications exclude 2010. * $p < 0.10$, ** $p < 0.05$, *** $p < 0.01$

**Table 5: Further robustness tests of the effect of ACA on health insurance among men in same-sex couples.**

|  | Control for state-year FE | Exclude states w/ SSM 2004-2009 | Exclude states w/ SSM 2004-2012 |
|---|---|---|---|
|  | (1) | (2) | (3) |
| *Any health insurance coverage* | | | |
| Age 21-25 * Post-2010 | 0.095** | 0.081** | 0.079** |
|  | (0.030) | (0.026) | (0.033) |
| N | 2,781 | 2,664 | 2,298 |
| Mean of DV for 21-25 pre-2010 | 0.627 | 0.621 | 0.610 |
| Adjusted R-squared | 0.189 | 0.175 | 0.175 |
| | | | |
| *Employer sponsored insurance* | | | |
| Age 21-25 * Post-2010 | 0.127** | 0.113** | 0.135*** |
|  | (0.045) | (0.037) | (0.039) |
| N | 2,781 | 2,664 | 2,298 |
| Mean of DV for 21-25 pre-2010 | 0.474 | 0.470 | 0.454 |
| Adjusted R-squared | 0.180 | 0.161 | 0.152 |
| | | | |
| *Controls for:* | | | |
| Age, state and year FE | X | X | X |
| State time-varying policies |  | X | X |
| Individual controls | X | X | X |
| State-year FE | X | | |

Notes: The dependent variable in the top panel is whether the respondent had any health insurance coverage. The dependent variable in the bottom panel is whether the respondent had health insurance through an employer. Sample includes male respondents in either married or unmarried same-sex couples. Individuals age 21-25 are compared to those age 27-31. The mean of the dependent variable only refers to individuals age 21-25 interviewed in 2008 or 2009. Same individual and state controls as Table 2. Column 1 includes state-year fixed effects. Column 2 excludes states that had legalized same-sex marriage between 2004 and 2009. Column 3 excludes states that had legalized same-sex marriage between 2004 and 2012. Standard errors clustered at the age level in parentheses. Weighted regressions using person weights. Source: ACS 2008-2012 (excluding 2010). * $p < 0.10$, ** $p < 0.05$, *** $p < 0.01$

**Table 6: Evidence on the mechanisms of the effect of ACA on health insurance among men in same-sex couples.**

|  | State of birth = current state of residence | State of birth ≠ current state of residence | Household heads | Spouses or partners |
|---|---|---|---|---|
|  | (1) | (2) | (3) | (4) |
| *Any health insurance coverage* | | | | |
| Age 21-25 * Post-2010 | 0.112** | 0.041 | 0.010 | 0.124*** |
|  | (0.038) | (0.058) | (0.043) | (0.037) |
| N | 1,266 | 1,150 | 1,235 | 1,546 |
| Mean of DV for 21-25 pre-2010 | 0.642 | 0.678 | 0.709 | 0.571 |
| Adjusted R-squared | 0.145 | 0.197 | 0.163 | 0.183 |
| | | | | |
| *Employer sponsored insurance* | | | | |
| Age 21-25 * Post-2010 | 0.204*** | 0.072 | -0.020 | 0.187*** |
|  | (0.035) | (0.083) | (0.069) | (0.049) |
| N | 1,266 | 1,150 | 1,235 | 1,546 |
| Mean of DV for 21-25 pre-2010 | 0.477 | 0.508 | 0.561 | 0.414 |
| Adjusted R-squared | 0.159 | 0.184 | 0.175 | 0.149 |
| | | | | |
| *Controls for:* | | | | |
| Age, State and year FE | X | X | X | X |
| State time-varying policies | X | X | X | X |
| Individual controls | X | X | X | X |

Notes: The dependent variable in the top panel is whether the respondent had any health insurance coverage. The dependent variable in the bottom panel is whether the respondent had health insurance through an employer. Sample includes male respondents in either married or unmarried same-sex couples. Individuals age 21-25 are compared to those age 27-31. The mean of the dependent variable only refers to individuals age 21-25 interviewed in 2008 or 2009. Same fixed effects, individual and state controls as Table 2. Column 1 includes only men whose current state of residence is the same of their state of birth. Column 2 includes only men whose current state of residence is different from their state of birth. Individuals born abroad have been excluded in columns 1-2. Column 3 includes only household heads. Column 4 includes only married spouses or unmarried partners. Standard errors clustered at the age level in parentheses. Weighted regressions using person weights. Source: ACS 2008-2012 (excluding 2010). * $p < 0.10$, ** $p < 0.05$, *** $p < 0.01$

**Table 7: Effect of ACA on additional outcomes for individuals in same-sex couples.**

|  | Employed | Unemployed | In the labor force | 40h/week or more | Number of h/week | Student |
|---|---|---|---|---|---|---|
|  | (1) | (2) | (3) | (4) | (5) | (6) |
| *Men* |  |  |  |  |  |  |
| Age 21-25 * Post-2010 | -0.016 | 0.013 | -0.003 | -0.016 | -0.502 | 0.003 |
|  | (0.035) | (0.021) | (0.034) | (0.051) | (1.761) | (0.025) |
| N | 2,781 | 2,781 | 2,781 | 2,781 | 2,781 | 2,781 |
| Mean of DV for 21-25 pre-2010 | 0.792 | 0.074 | 0.867 | 0.600 | 34.38 | 0.246 |
| Adjusted R-squared | 0.056 | 0.036 | 0.040 | 0.076 | 0.078 | 0.033 |
| *Women* |  |  |  |  |  |  |
| Age 21-25 * Post-2010 | -0.051 | 0.023* | -0.028 | -0.076** | -4.458*** | 0.019 |
|  | (0.032) | (0.012) | (0.031) | (0.029) | (0.563) | (0.031) |
| N | 3,614 | 3,614 | 3,614 | 3,614 | 3,614 | 3,614 |
| Mean of DV for 21-25 pre-2010 | 0.807 | 0.088 | 0.895 | 0.543 | 34.29 | 0.265 |
| Adjusted R-squared | 0.057 | 0.045 | 0.028 | 0.075 | 0.069 | 0.033 |
| *Controls for:* |  |  |  |  |  |  |
| Age, State and year FE | X | X | X | X | X | X |
| State time-varying policies | X | X | X | X | X | X |
| Individual controls | X | X | X | X | X | X |

Notes: The dependent variable is whether an individual was employed in column 1, whether an individual was unemployed in column 2, whether an individual was in the labor force in column 3, whether an individual usually worked at least 40h/week in column 4, number of hours usually worked per week in column 5, whether an individual was attending school in the three months preceding the interview in column 6. Sample includes male or female respondents in either married or unmarried same-sex couples. Individuals age 21-25 are compared to those age 27-31. The mean of the dependent variable only refers to individuals age 21-25 interviewed in 2008 or 2009. Same fixed effects, individual and state controls as Table 2. Standard errors clustered at the age level in parentheses. Weighted regressions using person weights. Source: ACS 2008-2012 (excluding 2010). * $p < 0.10$, ** $p < 0.05$, *** $p < 0.01$

# Online Appendix

## Appendix A: Variable description

### A.1 Dependent variables

*Any health insurance coverage* is an indicator equal to one if the respondent had any health insurance coverage at the time of the interview; zero otherwise. This includes employer-provided insurance, privately purchased insurance, Medicare, Medicaid or other governmental insurance, TRICARE or other military care or Veterans Administration-provided insurance. The Census Bureau does not consider respondents to have coverage if their only coverage is from Indian Health Services, as IHS policies are not always comprehensive.

*Employer-sponsored health insurance* is an indicator equal to one if the respondent had health insurance through a current employer, former employer, or union at the time of interview; zero otherwise. Importantly for our analysis, persons covered by another family member's current employer, former employer, or union are also coded as insured through an employer.

*Employed* is an indicator equal to one if the respondent worked at least one hour for pay or profit in the week preceding the interview, rather than being unemployed or not in the labor force. Unpaid family workers who worked at least 15 hours per week in the family business or farm are considered employed. On the other hand, housework at home is not included in this category. Respondents temporarily absent from their jobs (because of illness or vacation time) are still considered employed. Active military members are also coded as employed.

*Unemployed* is an indicator equal to one if the respondent did not have a job, was looking for a job, and had not yet found one at the time of the interview, rather than being employed or not in the labor force. Persons who had never worked but were actively seeking their first job are considered unemployed.

*In the labor force* is an indicator equal to one if the respondent was a part of the labor force, either working or seeking work, in the week preceding the interview; zero otherwise.

*Number of hours worked weekly.* The ACS reports the number of hours per week that the respondent usually worked, if the person worked during the 12 months preceding the interview. This variable is top coded at 99. Respondents who did not work in the 12 months preceding the interview are assigned value zero. From this variable we have generated the indicator *Working at least 40 hours per week* equal to one if the respondent used to work at least 40 hours per week; zero otherwise. Note that this variable is zero for respondents who did not work.

*Student status* is an indicator equal to one if the respondent attended school or college in the 3 months preceding the interview; zero otherwise.

### A.2 Individual-level controls

*Age* reports the respondent's age in years at the time of the interview.

*Race*. A series of indicator variables has been constructed to record the respondent's race: Black, Asian, or other races. Asian includes Chinese, Japanese, Other Asian or Pacific Islander. Other races include American Indian, Alaska Native, other race not listed, or individuals who selected two or three major races. White has been used as the comparison category.

*Hispanic* is an indicator equal to one if the respondent was identified as Mexican, Puerto Rican, Cuban, or Other Hispanic; zero otherwise.

*Higher Education* is an indicator equal to one if the respondent's highest degree completed was a Bachelor's degree or higher (Master's degree, Professional degree beyond a bachelor's degree, Doctoral degree); zero otherwise.

*Does not speak English* is an indicator equal to one if the respondent was not able to speak English; zero otherwise. This variable is self-reported.

## A.3 LGBT policy variables

*SSM legal* is an indicator variable equal to one in all states and time periods when same-sex marriage was legal; zero otherwise. The effective date has been used to code this variable. These data have been primarily obtained from the National Center for Lesbian Rights.[20]

*SSM ban* is a series of indicator variables equal to one in all states and time periods in which same-sex marriage was banned in the state constitution or state statute; zero otherwise. These indicators remain equal to one even in later years after the legalization of same-sex marriage in a given state. When more than one statutory ban was passed in a state, the oldest one has been used to code the state statute ban variable. These data have been primarily obtained from the Freedom to Marry campaign.[21]

*Domestic partnership* is an indicator variable equal to one in all states and time periods in which same-sex domestic partnerships were legal; zero otherwise. This indicator remains equal to one even in later years when\if a state had converted same-sex domestic partnerships into marriages. These data have been primarily obtained from the National Center for Lesbian Rights.[22]

*Civil union* is an indicator variable equal to one in all states and time periods in which same-sex civil unions were legal; zero otherwise. This indicator remains equal to one even in later years when\if a state had converted same-sex civil unions in marriages. These data have been primarily obtained from the National Center for Lesbian Rights.[23]

*Anti-discrimination law* is an indicator equal to one in all states and time periods in which employer discrimination based on sexual orientation was not allowed; zero otherwise. This variable has been set equal to one even if the law covered only sexual orientation, not gender identity, or if a law protecting trans individuals was passed at a later date. Laws protecting only public employees have not been considered. These data have been primarily obtained from the Freedom for All Americans campaign.[24]

---

[20] Source: http://www.nclrights.org/wp-content/uploads/2015/07/Relationship-Recognition.pdf. Accessed Oct/1/2019.
[21] Source: http://www.freedomtomarry.org/pages/winning-in-the-states. Accessed Oct/1/2019.
[22] See Footnote 5.
[23] See Footnote 5.

*Hate crime* is a series of indicator variables equal to one in all states and time periods in which there was a law specifically addressing hate or bias crimes based on sexual orientation only, or on sexual orientation and gender identity; zero otherwise. Since some states passed these laws after 2009, these variables have not been set equal to one for all states after President Obama signed the Matthew Shepard and James Byrd, Jr. Hate Crimes Prevention Act into law on October 28, 2009. These data have been primarily obtained from the Human Rights Campaign. [25]

**A.4 ACS state-level controls**

All of the ACS state level control variables have been computed using all individuals in the American Community Survey.

*Share black* reports for each year the proportion of state population that was black.

*Ethnic composition* reports for each year the proportion of state population that was Hispanic.

*Age 18-35* reports for each year the proportion of state population whose age was between 18 and 35.

*Proportion on welfare* reports for each year the proportion of state population that received income from various public assistance programs commonly referred to as "welfare". Assistance from private charities has not been included.

*Proportion unmarried* reports for each year the proportion of state different-sex couples (over all married and unmarried different-sex couples) that were unmarried.

**A.5 Additional state-level controls**

The following variables have been derived from data downloaded from the Bureau of Labor Statistics.[26]

*Population* records the estimates (in log) of the civilian noninstitutional population ages 16 and older computed by the Census Bureau.

*Unemployment rate* records the state-month unemployment rates for the civilian noninstitutional population ages 16 and older, not seasonally adjusted. From this, we have computed the average unemployment rate in each state.

---

[24] Source: https://www.freedomforallamericans.org/states/. Accessed: Oct/21/2019.
[25] Source: https://www.hrc.org/state-maps/hate-crimes. Accessed: Oct/25/2019.
[26] Source: https://www.bls.gov/lau/rdscnp16.htm. Accessed: Oct/1/2019.

*Income per capita* records the state-year personal income, not seasonally adjusted. The data have been retrieved from FRED, Federal Reserve Bank of St. Louis.[27]

### A.6 Additional policy controls

*ACA pre-expansion*. The Affordable Care Act (ACA) provided states with the option, effective April 2010, to receive federal Medicaid matching funds to cover low-income adults in order to get an early start on the 2014 Medicaid expansion. This indicator variable is equal to one in all states and time periods covered by an early Medicaid expansion to low-income adults through this new ACA option; zero otherwise. These data have been obtained from the Kaiser Family Foundation.[28]

*Medicaid expansion* is an indicator variable equal to one in all states and time periods covered by a 'regular' ACA Medicaid expansion (i.e., not a pre-expansion); zero otherwise. These data have been obtained from the Kaiser Family Foundation.[29]

*Private option* is an indicator variable equal to one in all states and time periods in which a state Medicaid program decided to buy private health insurance for its Medicaid population instead of providing coverage directly through the state's Medicaid program (or in which a private option waiver was effective); zero otherwise. These data have been obtained from Families USA.[30]

---

[27] Applied filters: income; not seasonally adjusted, per capita, state. Source: https://fred.stlouisfed.org/. Accessed: Oct/25/2019

[28] Source: https://www.kff.org/health-reform/issue-brief/states-getting-a-jump-start-on-health/. Accessed Oct/1/2019.

[29] Source: https://www.kff.org/medicaid/issue-brief/status-of-state-medicaid-expansion-decisions-interactive-map/. Accessed Oct/1/2019.

[30] Source: https://familiesusa.org/1115-waiver-element-private-option. Accessed Oct/1/2019.

# Appendix B: Additional figures and tables

## Table B1: Effect of ACA on single household heads.

|  | Single men | | Single women | |
|---|---|---|---|---|
|  | Any insurance | ESI | Any insurance | ESI |
|  | (1) | (2) | (3) | (4) |
| Age 21-25 * Post-2010 | 0.036*** | 0.059*** | 0.036*** | 0.053*** |
|  | (0.008) | (0.009) | (0.008) | (0.006) |
| N | 101,272 | 101,272 | 130,703 | 130,703 |
| Mean of DV for 21-25 pre-2010 | 0.711 | 0.532 | 0.757 | 0.460 |
| Adjusted R-squared | 0.120 | 0.089 | 0.080 | 0.157 |
| *Controls for:* | | | | |
| Age, State and year FE | X | X | X | X |
| State time-varying policies | X | X | X | X |
| Individual controls | X | X | X | X |

Notes: The dependent variable in columns 1 and 3 is whether the respondent had any health insurance coverage. The dependent variable in columns 2 and 4 is whether the respondent had health insurance through an employer. Sample includes male or female single household head respondents. Individuals age 21-25 are compared to those age 27-31. The mean of the dependent variable only refers to individuals age 21-25 interviewed in 2008 or 2009. Same fixed effects, individual and state controls as Table 2. Standard errors clustered at the age level in parentheses. Weighted regressions using person weights. Source: ACS 2008-2012 (excluding 2010). * $p < 0.10$, ** $p < 0.05$, *** $p < 0.01$

**Table B2: Effect of ACA on health insurance among same-sex couples. Placebo test.**

|  | Men | | Women | |
|---|---|---|---|---|
|  | Any insurance | ESI | Any insurance | ESI |
|  | (1) | (2) | (3) | (4) |
| Age 27-31 * Post-2010 | 0.010 | -0.009 | 0.019 | 0.010 |
|  | (0.031) | (0.038) | (0.023) | (0.028) |
| N | 3,857 | 3,857 | 4,680 | 4,680 |
| Mean of DV for 37-31 pre-2010 | 0.788 | 0.663 | 0.777 | 0.638 |
| Adjusted R-squared | 0.140 | 0.140 | 0.091 | 0.103 |
| *Controls for:* |  |  |  |  |
| Age, State and year FE | X | X | X | X |
| State time-varying policies | X | X | X | X |
| Individual controls | X | X | X | X |

Notes: The dependent variable in columns 1 and 3 is whether the respondent had any health insurance coverage. The dependent variable in columns 2 and 4 is whether the respondent had health insurance through an employer. Sample includes respondents in either married or unmarried different-sex or same-sex couples. Individuals age 27-31 are compared to those age 32-36. The mean of the dependent variable only refers to individuals age 27-31 interviewed in 2008 or 2009. Same fixed effects, individual and state controls as Table 2. Standard errors clustered at the age level in parentheses. Weighted regressions using person weights. Source: ACS 2008-2012 (excluding 2010). $^{*} p < 0.10$, $^{**} p < 0.05$, $^{***} p < 0.01$

**Table B3: Effect of ACA on health insurance among same-sex couples. Technical changes.**

|  | State-level clustered SE | Robust SE | Wild bootstrap | Effective cluster | Unweighted | Replication weights | Replication weights and age as PSU |
|---|---|---|---|---|---|---|---|
|  | (1) | (2) | (3) | (4) | (5) | (6) | (7) |
| *Men* | | | | | | | |
| Age 21-25 * Post-2010 | 0.080** | 0.080* | 0.080*** | 0.080*** | 0.081** | 0.080** | 0.080** |
|  | (0.029) | (0.056) | (0.005) | (0.009) | (0.010) | (0.049) | (0.049) |
| N sample | 2,781 | 2,781 | 2,781 | 2,781 | 2,781 | 2,781 | 2,781 |
| N population | -- | -- | -- | -- | -- | 290,925 | 290,925 |
| Mean of DV for 21-25 pre-2010 | 0.627 | 0.627 | 0.627 | 0.627 | 0.667 | 0.627 | 0.627 |
| Adjusted R-squared | 0.176 | 0.176 | 0.176 | 0.176 | 0.132 | 0.201 | 0.201 |
| | | | | | | | |
| *Women* | | | | | | | |
| Age 21-25 * Post-2010 | 0.015 | 0.015 | 0.015 | 0.015 | 0.007 | 0.015 | 0.015 |
|  | (0.677) | (0.688) | (0.709) | (0.705) | (0.792) | (0.713) | (0.713) |
| N sample | 3,614 | 3,614 | 3,614 | 3,614 | 3,614 | 3,614 | 3,614 |
| N population | -- | -- | -- | -- | -- | 367,445 | 367,445 |
| Mean of DV for 21-25 pre-2010 | 0.659 | 0.659 | 0.659 | 0.659 | 0.683 | 0.659 | 0.659 |
| Adjusted R-squared | 0.121 | 0.121 | 0.121 | 0.121 | 0.101 | 0.142 | 0.142 |
| | | | | | | | |
| *Controls for:* | | | | | | | |
| Age, State and year FE | X | X | X | X | X | X | X |
| State time-varying controls | X | X | X | X | X | X | X |
| Individual controls | X | X | X | X | X | X | X |

Notes: The dependent variable is whether the respondent had any health insurance coverage. Sample includes respondents in either married or unmarried same-sex couples. Individuals age 21-25 are compared to those age 27-31. P-values shown in parenthesis instead of standard errors. Standard errors in column 1 are clustered at the state level. Standard errors in column 2 are not clustered. Wild bootstrapped p-values in column 3 have been computed using the command boottest in Stata (Roodman et al. 2019). P-values in columns 4 have been computed from the effective number of clusters using the command clusteff in Stata (Lee and Steigerwald 2018). The estimates in column 5 are unweighted. The estimates in columns 6-7 have been obtained using replication weights as described by IPUMS (https://usa.ipums.org/usa/repwt.shtml). Same fixed effects, individual and state controls as Table 2. Source: ACS 2008-2012 (excluding 2010). * $p < 0.10$, ** $p < 0.05$, *** $p < 0.01$

**Table B4: Effect of ACA on health insurance among same-sex couples. Additional extensions.**

|  | Only unmarried | Include 2010 as treated | Include 2010 as control | Include 26 as control | 23-25 vs 27-29 |
|---|---|---|---|---|---|
|  | (1) | (2) | (3) | (4) | (5) |
| *Men* |  |  |  |  |  |
| Treated age group * Post-2010 | 0.095*** | 0.074** | 0.067*** | 0.085*** | 0.061** |
|  | (0.022) | (0.030) | (0.018) | (0.023) | (0.021) |
| N | 2,670 | 3,471 | 3,471 | 3,087 | 1,739 |
| Mean of DV for 21-25 pre-2010 | 0.627 | 0.627 | 0.625 | 0.627 | 0.653 |
| Adjusted R-squared | 0.176 | 0.155 | 0.155 | 0.172 | 0.178 |
|  |  |  |  |  |  |
| *Women* |  |  |  |  |  |
| Treated age group * Post-2010 | 0.013 | -0.005 | 0.034 | 0.023 | 0.017 |
|  | (0.038) | (0.027) | (0.037) | (0.037) | (0.050) |
| N | 3,414 | 4,537 | 4,537 | 3,998 | 2,228 |
| Mean of DV for 21-25 pre-2010 | 0.659 | 0.659 | 0.646 | 0.659 | 0.688 |
| Adjusted R-squared | 0.122 | 0.120 | 0.121 | 0.117 | 0.122 |
|  |  |  |  |  |  |
| *Controls for:* |  |  |  |  |  |
| Age, State and year FE | X | X | X | X | X |
| State time-varying controls | X | X | X | X | X |
| Individual controls | X | X | X | X | X |

Notes: The dependent variable is whether the respondent had any health insurance coverage. Column 1 excludes married same-sex couples (identified in 2012). Column 2 includes young respondents (21-25) from 2010 in the treatment group, older respondent (27-31) in the control group. Column 2 includes all respondents from 2010 in the control group. Column 4 includes respondents age 26 in the control group (age 26-31). Column 5 compares individuals age 23-25 to those age 27-29. Same fixed effect, individual and state controls as Table 2. Standard errors clustered at the age level in parentheses. Weighted regressions using person weights. Source: ACS 2008-2012 (excluding 2010, except in columns 2-3). * $p < 0.10$, ** $p < 0.05$, *** $p < 0.01$

**Table B5: Effect of ACA on health insurance among men in SSC. Exclude one state at a time**

| Excluded state | Any health insurance coverage | | Employer sponsored insurance | |
|---|---|---|---|---|
| Alabama | 0.081*** | (0.023) | 0.120** | (0.039) |
| Alaska | 0.079*** | (0.023) | 0.110** | (0.040) |
| Arizona | 0.077*** | (0.021) | 0.106** | (0.038) |
| Arkansas | 0.079*** | (0.023) | 0.112** | (0.040) |
| California | 0.075** | (0.025) | 0.112** | (0.043) |
| Colorado | 0.087** | (0.027) | 0.117** | (0.045) |
| Connecticut | 0.076** | (0.024) | 0.106** | (0.040) |
| Delaware | 0.081*** | (0.023) | 0.112** | (0.040) |
| DC | 0.083*** | (0.023) | 0.119** | (0.040) |
| Florida | 0.076** | (0.024) | 0.094* | (0.044) |
| Georgia | 0.093*** | (0.022) | 0.123** | (0.040) |
| Hawaii | 0.079*** | (0.023) | 0.113** | (0.039) |
| Idaho | 0.080*** | (0.022) | 0.111** | (0.039) |
| Illinois | 0.064** | (0.025) | 0.092* | (0.042) |
| Indiana | 0.075*** | (0.021) | 0.106** | (0.042) |
| Iowa | 0.078*** | (0.022) | 0.107** | (0.039) |
| Kansas | 0.081*** | (0.024) | 0.113** | (0.040) |
| Kentucky | 0.084*** | (0.023) | 0.113** | (0.040) |
| Louisiana | 0.083*** | (0.023) | 0.114** | (0.038) |
| Maine | 0.077*** | (0.024) | 0.110** | (0.039) |
| Maryland | 0.083*** | (0.023) | 0.114** | (0.039) |
| Massachusetts | 0.087*** | (0.025) | 0.123*** | (0.037) |
| Michigan | 0.077*** | (0.022) | 0.114** | (0.041) |
| Minnesota | 0.079*** | (0.024) | 0.115** | (0.036) |
| Mississippi | 0.081*** | (0.023) | 0.113** | (0.040) |
| Missouri | 0.079*** | (0.023) | 0.109** | (0.040) |
| Montana | 0.082*** | (0.023) | 0.115** | (0.039) |
| Nebraska | 0.080*** | (0.023) | 0.111** | (0.040) |
| Nevada | 0.077*** | (0.021) | 0.108** | (0.038) |
| New Hampshire | 0.081*** | (0.022) | 0.113** | (0.038) |
| New Jersey | 0.078*** | (0.023) | 0.116** | (0.038) |
| New Mexico | 0.071*** | (0.020) | 0.103** | (0.035) |
| New York | 0.088*** | (0.026) | 0.129** | (0.043) |
| North Carolina | 0.081** | (0.028) | 0.103* | (0.047) |
| North Dakota | 0.080*** | (0.023) | 0.111** | (0.040) |
| Ohio | 0.076** | (0.029) | 0.098* | (0.044) |
| Oklahoma | 0.083*** | (0.023) | 0.115** | (0.041) |
| Oregon | 0.088*** | (0.025) | 0.113** | (0.040) |
| Pennsylvania | 0.088*** | (0.023) | 0.114** | (0.041) |
| Rhode Island | 0.077*** | (0.023) | 0.108** | (0.039) |
| South Carolina | 0.081*** | (0.023) | 0.111** | (0.040) |
| South Dakota | 0.079*** | (0.022) | 0.110** | (0.039) |
| Tennessee | 0.087*** | (0.023) | 0.115** | (0.038) |
| Texas | 0.080** | (0.030) | 0.114** | (0.045) |
| Utah | 0.079*** | (0.023) | 0.116** | (0.039) |
| Vermont | 0.080*** | (0.023) | 0.111** | (0.040) |
| Virginia | 0.090*** | (0.025) | 0.109** | (0.043) |
| Washington | 0.067** | (0.024) | 0.106** | (0.035) |
| West Virginia | 0.080*** | (0.023) | 0.111** | (0.040) |
| Wisconsin | 0.084** | (0.026) | 0.118** | (0.040) |
| Wyoming | 0.078*** | (0.023) | 0.110** | (0.040) |

Reported coefficient of age 21-25*post-2010. Same structure as Column 1-2 Table 2. $^{*} p < 0.10$, $^{**} p < 0.05$, $^{***} p < 0.01$

**Table B6: Effect of ACA on health insurance among men in same-sex couples. By education.**

|  | All | High school or less | Some college | BA or more |
|---|---|---|---|---|
|  | (1) | (2) | (3) | (4) |
| *Any health insurance coverage* | | | | |
| Age 21-25 * Post-2010 | 0.080*** | 0.177* | 0.060 | 0.036 |
|  | (0.023) | (0.079) | (0.055) | (0.064) |
| N | 2,781 | 659 | 1,038 | 1,084 |
| Mean of DV for 21-25 pre-2010 | 0.627 | 0.454 | 0.640 | 0.880 |
| Adjusted R-squared | 0.176 | 0.101 | 0.099 | 0.078 |
| | | | | |
| *Employer sponsored insurance* | | | | |
| Age 21-25 * Post-2010 | 0.111** | 0.236*** | 0.128** | 0.052 |
|  | (0.040) | (0.060) | (0.055) | (0.065) |
| N | 2,781 | 659 | 1,038 | 1,084 |
| Mean of DV for 21-25 pre-2010 | 0.474 | 0.270 | 0.493 | 0.766 |
| Adjusted R-squared | 0.161 | 0.091 | 0.085 | 0.064 |
| | | | | |
| *Controls for:* | | | | |
| Age, State and year FE | X | X | X | X |
| State time-varying policies | X | X | X | X |
| Individual controls | X | X | X | X |

Notes: The dependent variable in the top panel is whether the respondent had any health insurance coverage. The dependent variable in the bottom panel is whether the respondent had health insurance through an employer. Sample includes male respondents in either married or unmarried same-sex couples. Individuals age 21-25 are compared to those age 27-31. The mean of the dependent variable only refers to individuals age 21-25 interviewed in 2008 or 2009. Same individual and state controls as Table 2. Column 1 includes all individuals. Column 2 includes only individuals with a high school degree, GED, or less than high school. Column 3 includes only individuals with some college education or an associate degree. Column 4 includes individuals with a bachelor's degree or a higher educational level. Standard errors clustered at the age level in parentheses. Weighted regressions using person weights. Source: ACS 2008-2012 (excluding 2010). * $p < 0.10$, ** $p < 0.05$, *** $p < 0.01$

**Table B7: Effect of ACA on health insurance among same-sex couples. By race.**

|  | All | White | Black | Other | Hispanic |
|---|---|---|---|---|---|
|  | (1) | (2) | (3) | (4) | (5) |
| *Men* |  |  |  |  |  |
| Age 21-25 * Post-2010 | 0.080*** | 0.113*** | -0.166 | 0.044 | 0.106 |
|  | (0.023) | (0.023) | (0.178) | (0.102) | (0.096) |
| N | 2,781 | 2,222 | 194 | 365 | 480 |
| Mean of DV for 21-25 pre-2010 | 0.627 | 0.653 | 0.530 | 0.510 | 0.399 |
| Adjusted R-squared | 0.176 | 0.153 | 0.111 | 0.215 | 0.224 |
|  |  |  |  |  |  |
| *Employer sponsored insurance* |  |  |  |  |  |
| Age 21-25 * Post-2010 | 0.111** | 0.139** | 0.130 | -0.023 | 0.099 |
|  | (0.040) | (0.045) | (0.230) | (0.083) | (0.114) |
| N | 2781 | 2,222 | 194 | 365 | 480 |
| Mean of DV for 21-25 pre-2010 | 0.474 | 0.497 | 0.261 | 0.439 | 0.323 |
| Adjusted R-squared | 0.161 | 0.137 | 0.194 | 0.218 | 0.185 |
|  |  |  |  |  |  |
| *Controls for:* |  |  |  |  |  |
| Age, State and year FE | X | X | X | X | X |
| State time-varying controls | X | X | X | X | X |
| Individual controls | X | X | X | X | X |

Notes: The dependent variable in the top panel is whether the respondent had any health insurance coverage. The dependent variable in the bottom panel is whether the respondent had health insurance through an employer. Sample includes male respondents in either married or unmarried same-sex couples. Individuals age 21-25 are compared to those age 27-31. The mean of the dependent variable only refers to individuals age 21-25 interviewed in 2008 or 2009. Column 4 includes Asian, Pacific Islander, American Indian, Alaska Native, other or mixed races. Same fixed effect and state controls as Table 2. Individual controls: language and education. Standard errors clustered at the age level in parentheses. Weighted regressions using person weights. Source: ACS 2008-2012 (excluding 2010). * $p < 0.10$, ** $p < 0.05$, *** $p < 0.01$

**Table B8: Evidence on the mechanisms of the effect of ACA on health insurance among women in same-sex couples.**

|  | State of birth = current state of residence | State of birth ≠ current state of residence | Household heads | Spouses or partners |
|---|---|---|---|---|
|  | (1) | (2) | (3) | (4) |
| *Any health insurance coverage* |  |  |  |  |
| Age 21-25 * Post-2010 | 0.088* | -0.012 | 0.015 | 0.020 |
|  | (0.047) | (0.076) | (0.049) | (0.057) |
| N | 1,929 | 1,422 | 1,687 | 1,927 |
| Mean of DV for 21-25 pre-2010 | 0.656 | 0.682 | 0.679 | 0.642 |
| Adjusted R-squared | 0.119 | 0.142 | 0.113 | 0.127 |
|  |  |  |  |  |
| *Employer sponsored insurance* |  |  |  |  |
| Age 21-25 * Post-2010 | 0.086 | 0.026 | -0.030 | 0.095** |
|  | (0.057) | (0.058) | (0.061) | (0.041) |
| N | 1,929 | 1,422 | 1,687 | 1,927 |
| Mean of DV for 21-25 pre-2010 | 0.476 | 0.482 | 0.497 | 0.456 |
| Adjusted R-squared | 0.140 | 0.128 | 0.119 | 0.128 |
|  |  |  |  |  |
| *Controls for:* |  |  |  |  |
| Age, State and year FE | X | X | X | X |
| State time-varying policies | X | X | X | X |
| Individual controls | X | X | X | X |

Notes: The dependent variable in the top panel is whether the respondent had any health insurance coverage. The dependent variable in the bottom panel is whether the respondent had health insurance through an employer. Sample includes female respondents in either married or unmarried same-sex couples. Individuals age 21-25 are compared to those age 27-31. The mean of the dependent variable only refers to individuals age 21-25 interviewed in 2008 or 2009. Same fixed effects, individual and state controls as Table 2. Column 1 includes only women whose current state of residence is the same of their state of birth. Column 2 includes only women whose current state of residence is different from their state of birth. Individuals born abroad have been excluded in columns 1-2. Column 3 includes only household heads. Column 4 includes only married spouses or unmarried partners. Standard errors clustered at the age level in parentheses. Weighted regressions using person weights. Source: ACS 2008-2012 (excluding 2010). * $p < 0.10$, ** $p < 0.05$, *** $p < 0.01$